%% file: main.tex
\documentclass{class/llncs}

\usepackage{epsfig}
\usepackage{fancyhdr}
\usepackage{plain}

\usepackage{footmisc}
\usepackage[FIGBOTCAP,TABBOTCAP,center,nooneline]{subfigure}
\usepackage[dvips]{rotating}
\usepackage[noend]{algorithmic}
\usepackage[normalem]{ulem}
\usepackage{amscd}
\usepackage{amsfonts}
\usepackage{amsmath}
\usepackage{booktabs}
\usepackage{color}
\usepackage{comment}
\usepackage{datetime}
\usepackage{epsfig}
\usepackage{graphicx}
\usepackage{pdfpages}
\usepackage{latexsym}
\usepackage{marginnote}
\usepackage{mathtools}
\usepackage{multirow}
\usepackage{paralist}
\usepackage{placeins}
\usepackage{rotate}
\usepackage{tikz}
\usepackage{times}
\usepackage{url}
\usepackage{varwidth}
\usepackage{verbatim}
\usepackage{wrapfig}
\usepackage{xcolor,colortbl}
\usepackage{transparent}
\usepackage{xspace}
\usepackage[formats]{listings}
\usepackage{courier}
\usepackage{relsize}
\usepackage{float}
\usepackage{afterpage}
\usepackage{tabularx}
\usepackage[linesnumbered,ruled,vlined]{algorithm2e}
\usepackage[framemethod=tikz]{mdframed}
\usepackage[bookmarks=false,colorlinks]{hyperref}
\usepackage{cleveref}
\usepackage{etoolbox}
\usepackage{array}
\usepackage{eqparbox}
\usepackage[T1]{fontenc}
\usepackage[utf8]{inputenc}
\usepackage{bibentry}
\usepackage{nicefrac}
\usepackage{xfrac}

\input{options.tex}

\input{macros/rs_macros_colors}

\input{macros/rs_macros_dl}
\input{macros/rs_macros_general}


\input{macros/rs_macros_optsmt}

\input{macros/rs_macros_sat}

\input{macros/rs_macros_smt}

\input{macros/rs_macros_long}

\input{macros/pt_macros_abrv}

\input{macros/pt_macros_notes}
\input{macros/pt_macros_lst}
\input{macros/pt_macros_cleveref}
\input{macros/pt_macros_alg}

\input{macros/pt_macros_proofs}

\input{macros/rs_macros_specific}
\renewcommand{\marg}[1]{}
\newcommand{\ignoreforblind}[1]{}

\input{macros/macros_anon}

\input{macros/fc_macros_notes}

\begin{document}

\pagestyle{plain}
\pagenumbering{arabic}

\title{From \minizinc{} to Optimization Modulo Theories, \\and Back (Extended Version)}
\author{Francesco Contaldo, Patrick Trentin, Roberto Sebastiani }
\institute{DISI, University of Trento, Italy}

\maketitle

\ignoreinshort{\large\begin{center}\noi\textcolor{blue}{(long version of CPAIOR 2020 publication)}\end{center}}

\begin{abstract}
\input{src/abstract.tex}

\end{abstract}

\section{Introduction}
\label{sec:intro}
\input{src/introduction.tex}

\section{Background}
\label{sec:background}

    \label{sec:background_omt}
    \input{src/background_omt.tex}

    \label{sec:background_mzn}
    \input{src/background_mzn.tex}

\section{From \minizinc{} to \omt{}}
\label{sec:mzn2omt}
\input{src/mzn2omt/introduction.tex}

    \label{sec:mzn2omt_challenges}
    \input{src/mzn2omt/challenges.tex}

    \label{sec:mzn2omt_solutions}
    \input{src/mzn2omt/solutions.tex}

\section{From \omt{} to \minizinc{}}
\label{sec:omt2mzn}
\input{src/omt2mzn/introduction.tex}

    \label{sec:omt2mzn_challenges}
    \input{src/omt2mzn/challenges.tex}

\section{Experimental Evaluations}
\label{sec:expeval}
\input{src/expeval.tex}

    \subsection{Evaluation on \minizinc{} Benchmark Sets}
    \label{sec:mzn2omt_expeval}
    \input{src/mzn2omt/expeval.tex}

    \subsection{Evaluation on \omt{} Benchmark Sets}
    \label{sec:omt2mzn_expeval}
    \input{src/omt2mzn/expeval.tex}

    \subsection{Discussion}
    \label{sec:discussion}
    \input{src/discussion_features.tex}

\section{Conclusions \& Future Work}
\label{sec:conclusions}
\input{src/conclusions.tex}


\newpage
\FloatBarrier
\pagenumbering{roman}
\bibliographystyle{abbrv}
\bibliography{bibs/rs_refs,bibs/rs_ownrefs,bibs/rs_specific,bibs/sathandbook,bibs/pt_refs,bibs/fc_refs,bibs/pt_urls,bibs/pt_sat_refs,bibs/pt_cp,bibs/omt_refs}


\end{document}

%% file: options.tex
\newcommand{\longversion}{true}
\newcommand{\withproofs}{false}

%% file: macros/rs_macros_colors.tex
\newcommand{\blue}[1]{\textcolor{blue}{#1}}

\newcommand{\red}[1]{\textcolor{red}{#1}}
\newcommand{\green}[1]{\textcolor{darkgreen}{#1}}

%% file: macros/rs_macros_general.tex
\input{macros/rgb}

\newcommand{\TODO}[1]{{}}

\newcommand{\ignore}[1]{}
\newcommand{\todo}[1] {}

\newcommand{\RSTODO}[1]{{\bf \textcolor{darkgreen}{{\fbox{RS TODO:} #1}}}}

\newcommand{\marg}[1]{\marginpar{\ \\{\em #1\/}}}
\renewcommand{\marg}[1]{\marginpar{\ \\\textcolor{blue}{\ {\sf #1\/}}}}

\newcommand{\RSNOTE}[1]{\marginpar{\textcolor{darkgreen}{\textbf{RS: }
      }}}
\renewcommand{\RSNOTE}[1]{\textcolor{darkgreen}{{\bf RS: {\footnotesize #1}}}}

 \newcommand{\ignoreinshort}[1]{}
 \newcommand{\ignoreinlong}[1]{{#1}}

\def\makenewenumerate#1#2{%
\newcounter{cnt#1}
\newenvironment{#1}%
{\begin{list}{\makebox[0pt][r]{#2}}%
{\setlength{\itemsep}{0pt}%
 \setlength{\parsep}{.2em}%
 \setlength{\leftmargin}{1.5em}%
 \setlength{\labelwidth}{.4em}%
 \usecounter{cnt#1}}}
{\end{list}}}

\makenewenumerate{myenumerate}{\arabic{cntmyenumerate}.}

\makenewenumerate{renumerate}{\rm(\roman{cntrenumerate})}
\makenewenumerate{renumerateprime}{\rm(\roman{cntrenumerateprime}$'$)}
\makenewenumerate{renumeratesecond}{\rm(\roman{cntrenumeratesecond}$''$)}

\makenewenumerate{aenumerate}{({\it\alph{cntaenumerate}})}
\makenewenumerate{aenumerateprime}{({\it\alph{cntaenumerateprime}$'$})}
\makenewenumerate{aenumeratesecond}{({\it\alph{cntaenumeratesecond}$''$})}

\def\newplaintheorem#1#2{%
\newtheorem{#1plain}{#2}[section]%
\newenvironment{#1}{\begin{#1plain}\rm }{\end{#1plain}}}

\newplaintheorem{notation}{Notation}
\newplaintheorem{cexample}{Counter-Example}

\newcommand{\sref}[1]{\S{}\ref{#1}}
\newcommand{\noi}{\noindent}

\newcommand{\tuple}[1]{\ensuremath{\langle{#1}\rangle}\xspace}
\newcommand{\set}[1]{\ensuremath{\{{#1}\}}\xspace}

\newcommand{\defas}{\ensuremath{\stackrel{\text{\tiny def}}{=}}\xspace}

\newcommand\cala{\ensuremath{\mathcal{A}}\xspace}



%% file: macros/rgb.tex
\definecolor {snow}                {rgb}{1.00,0.98,0.98}
\definecolor {ghostwhite}          {rgb}{0.97,0.97,1.00}
\definecolor {whitesmoke}          {rgb}{0.96,0.96,0.96}
\definecolor {gainsboro}           {rgb}{0.86,0.86,0.86}
\definecolor {floralwhite}         {rgb}{1.00,0.98,0.94}
\definecolor {oldlace}             {rgb}{0.99,0.96,0.90}
\definecolor {linen}               {rgb}{0.98,0.94,0.90}
\definecolor {antiquewhite}        {rgb}{0.98,0.92,0.84}
\definecolor {papayawhip}          {rgb}{1.00,0.94,0.84}
\definecolor {blanchedalmond}      {rgb}{1.00,0.92,0.80}
\definecolor {bisque}              {rgb}{1.00,0.89,0.77}
\definecolor {peachpuff}           {rgb}{1.00,0.85,0.73}
\definecolor {navajowhite}         {rgb}{1.00,0.87,0.68}
\definecolor {moccasin}            {rgb}{1.00,0.89,0.71}
\definecolor {cornsilk}            {rgb}{1.00,0.97,0.86}
\definecolor {ivory}               {rgb}{1.00,1.00,0.94}
\definecolor {lemonchiffon}        {rgb}{1.00,0.98,0.80}
\definecolor {seashell}            {rgb}{1.00,0.96,0.93}
\definecolor {honeydew}            {rgb}{0.94,1.00,0.94}
\definecolor {mintcream}           {rgb}{0.96,1.00,0.98}
\definecolor {azure}               {rgb}{0.94,1.00,1.00}
\definecolor {aliceblue}           {rgb}{0.94,0.97,1.00}
\definecolor {lavender}            {rgb}{0.90,0.90,0.98}
\definecolor {lavenderblush}       {rgb}{1.00,0.94,0.96}
\definecolor {mistyrose}           {rgb}{1.00,0.89,0.88}
\definecolor {white}               {rgb}{1.00,1.00,1.00}
\definecolor {black}               {rgb}{0.00,0.00,0.00}
\definecolor {darkslategray}       {rgb}{0.18,0.31,0.31}
\definecolor {dimgray}             {rgb}{0.41,0.41,0.41}
\definecolor {slategray}           {rgb}{0.44,0.50,0.56}
\definecolor {lightslategray}      {rgb}{0.47,0.53,0.60}
\definecolor {gray}                {rgb}{0.75,0.75,0.75}
\definecolor {lightgrey}           {rgb}{0.83,0.83,0.83}
\definecolor {midnightblue}        {rgb}{0.10,0.10,0.44}
\definecolor {navy}                {rgb}{0.00,0.00,0.50}
\definecolor {cornflowerblue}      {rgb}{0.39,0.58,0.93}
\definecolor {darkslateblue}       {rgb}{0.28,0.24,0.55}
\definecolor {slateblue}           {rgb}{0.42,0.35,0.80}
\definecolor {mediumslateblue}     {rgb}{0.48,0.41,0.93}
\definecolor {lightslateblue}      {rgb}{0.52,0.44,1.00}
\definecolor {mediumblue}          {rgb}{0.00,0.00,0.80}
\definecolor {royalblue}           {rgb}{0.25,0.41,0.88}
\definecolor {blue}                {rgb}{0.00,0.00,1.00}
\definecolor {dodgerblue}          {rgb}{0.12,0.56,1.00}
\definecolor {deepskyblue}         {rgb}{0.00,0.75,1.00}
\definecolor {skyblue}             {rgb}{0.53,0.81,0.92}
\definecolor {lightskyblue}        {rgb}{0.53,0.81,0.98}
\definecolor {steelblue}           {rgb}{0.27,0.51,0.71}
\definecolor {lightsteelblue}      {rgb}{0.69,0.77,0.87}
\definecolor {lightblue}           {rgb}{0.68,0.85,0.90}
\definecolor {powderblue}          {rgb}{0.69,0.88,0.90}
\definecolor {paleturquoise}       {rgb}{0.69,0.93,0.93}
\definecolor {darkturquoise}       {rgb}{0.00,0.81,0.82}
\definecolor {mediumturquoise}     {rgb}{0.28,0.82,0.80}
\definecolor {turquoise}           {rgb}{0.25,0.88,0.82}
\definecolor {cyan}                {rgb}{0.00,1.00,1.00}
\definecolor {lightcyan}           {rgb}{0.88,1.00,1.00}
\definecolor {cadetblue}           {rgb}{0.37,0.62,0.63}
\definecolor {mediumaquamarine}    {rgb}{0.40,0.80,0.67}
\definecolor {aquamarine}          {rgb}{0.50,1.00,0.83}
\definecolor {darkgreen}           {rgb}{0.00,0.39,0.00}
\definecolor {darkolivegreen}      {rgb}{0.33,0.42,0.18}
\definecolor {darkseagreen}        {rgb}{0.56,0.74,0.56}
\definecolor {seagreen}            {rgb}{0.18,0.55,0.34}
\definecolor {mediumseagreen}      {rgb}{0.24,0.70,0.44}
\definecolor {lightseagreen}       {rgb}{0.13,0.70,0.67}
\definecolor {palegreen}           {rgb}{0.60,0.98,0.60}
\definecolor {springgreen}         {rgb}{0.00,1.00,0.50}
\definecolor {lawngreen}           {rgb}{0.49,0.99,0.00}
\definecolor {green}               {rgb}{0.00,1.00,0.00}
\definecolor {chartreuse}          {rgb}{0.50,1.00,0.00}
\definecolor {mediumspringgreen}   {rgb}{0.00,0.98,0.60}
\definecolor {greenyellow}         {rgb}{0.68,1.00,0.18}
\definecolor {limegreen}           {rgb}{0.20,0.80,0.20}
\definecolor {yellowgreen}         {rgb}{0.60,0.80,0.20}
\definecolor {forestgreen}         {rgb}{0.13,0.55,0.13}
\definecolor {olivedrab}           {rgb}{0.42,0.56,0.14}
\definecolor {darkkhaki}           {rgb}{0.74,0.72,0.42}
\definecolor {khaki}               {rgb}{0.94,0.90,0.55}
\definecolor {palegoldenrod}       {rgb}{0.93,0.91,0.67}
\definecolor {lightgoldenrodyellow} {rgb}{0.98,0.98,0.82}
\definecolor {lightyellow}         {rgb}{1.00,1.00,0.88}
\definecolor {yellow}              {rgb}{1.00,1.00,0.00}
\definecolor {gold}                {rgb}{1.00,0.84,0.00}
\definecolor {lightgoldenrod}      {rgb}{0.93,0.87,0.51}
\definecolor {goldenrod}           {rgb}{0.85,0.65,0.13}
\definecolor {darkgoldenrod}       {rgb}{0.72,0.53,0.04}
\definecolor {rosybrown}           {rgb}{0.74,0.56,0.56}
\definecolor {indianred}           {rgb}{0.80,0.36,0.36}
\definecolor {saddlebrown}         {rgb}{0.55,0.27,0.07}
\definecolor {sienna}              {rgb}{0.63,0.32,0.18}
\definecolor {peru}                {rgb}{0.80,0.52,0.25}
\definecolor {burlywood}           {rgb}{0.87,0.72,0.53}
\definecolor {beige}               {rgb}{0.96,0.96,0.86}
\definecolor {wheat}               {rgb}{0.96,0.87,0.70}
\definecolor {sandybrown}          {rgb}{0.96,0.64,0.38}
\definecolor {tan}                 {rgb}{0.82,0.71,0.55}
\definecolor {chocolate}           {rgb}{0.82,0.41,0.12}
\definecolor {firebrick}           {rgb}{0.70,0.13,0.13}
\definecolor {brown}               {rgb}{0.65,0.16,0.16}
\definecolor {darksalmon}          {rgb}{0.91,0.59,0.48}
\definecolor {salmon}              {rgb}{0.98,0.50,0.45}
\definecolor {lightsalmon}         {rgb}{1.00,0.63,0.48}
\definecolor {orange}              {rgb}{1.00,0.65,0.00}
\definecolor {darkorange}          {rgb}{1.00,0.55,0.00}
\definecolor {coral}               {rgb}{1.00,0.50,0.31}
\definecolor {lightcoral}          {rgb}{0.94,0.50,0.50}
\definecolor {tomato}              {rgb}{1.00,0.39,0.28}
\definecolor {orangered}           {rgb}{1.00,0.27,0.00}
\definecolor {red}                 {rgb}{1.00,0.00,0.00}
\definecolor {hotpink}             {rgb}{1.00,0.41,0.71}
\definecolor {deeppink}            {rgb}{1.00,0.08,0.58}
\definecolor {pink}                {rgb}{1.00,0.75,0.80}
\definecolor {lightpink}           {rgb}{1.00,0.71,0.76}
\definecolor {palevioletred}       {rgb}{0.86,0.44,0.58}
\definecolor {maroon}              {rgb}{0.69,0.19,0.38}
\definecolor {mediumvioletred}     {rgb}{0.78,0.08,0.52}
\definecolor {violetred}           {rgb}{0.82,0.13,0.56}
\definecolor {magenta}             {rgb}{1.00,0.00,1.00}
\definecolor {violet}              {rgb}{0.93,0.51,0.93}
\definecolor {plum}                {rgb}{0.87,0.63,0.87}
\definecolor {orchid}              {rgb}{0.85,0.44,0.84}
\definecolor {mediumorchid}        {rgb}{0.73,0.33,0.83}
\definecolor {darkorchid}          {rgb}{0.60,0.20,0.80}
\definecolor {darkviolet}          {rgb}{0.58,0.00,0.83}
\definecolor {blueviolet}          {rgb}{0.54,0.17,0.89}
\definecolor {purple}              {rgb}{0.63,0.13,0.94}
\definecolor {mediumpurple}        {rgb}{0.58,0.44,0.86}
\definecolor {thistle}             {rgb}{0.85,0.75,0.85}
\definecolor {snow2}               {rgb}{0.93,0.91,0.91}
\definecolor {snow3}               {rgb}{0.80,0.79,0.79}
\definecolor {snow4}               {rgb}{0.55,0.54,0.54}
\definecolor {seashell2}           {rgb}{0.93,0.90,0.87}
\definecolor {seashell3}           {rgb}{0.80,0.77,0.75}
\definecolor {seashell4}           {rgb}{0.55,0.53,0.51}
\definecolor {antiquewhite1}       {rgb}{1.00,0.94,0.86}
\definecolor {antiquewhite2}       {rgb}{0.93,0.87,0.80}
\definecolor {antiquewhite3}       {rgb}{0.80,0.75,0.69}
\definecolor {antiquewhite4}       {rgb}{0.55,0.51,0.47}
\definecolor {bisque2}             {rgb}{0.93,0.84,0.72}
\definecolor {bisque3}             {rgb}{0.80,0.72,0.62}
\definecolor {bisque4}             {rgb}{0.55,0.49,0.42}
\definecolor {peachpuff2}          {rgb}{0.93,0.80,0.68}
\definecolor {peachpuff3}          {rgb}{0.80,0.69,0.58}
\definecolor {peachpuff4}          {rgb}{0.55,0.47,0.40}
\definecolor {navajowhite2}        {rgb}{0.93,0.81,0.63}
\definecolor {navajowhite3}        {rgb}{0.80,0.70,0.55}
\definecolor {navajowhite4}        {rgb}{0.55,0.47,0.37}
\definecolor {lemonchiffon2}       {rgb}{0.93,0.91,0.75}
\definecolor {lemonchiffon3}       {rgb}{0.80,0.79,0.65}
\definecolor {lemonchiffon4}       {rgb}{0.55,0.54,0.44}
\definecolor {cornsilk2}           {rgb}{0.93,0.91,0.80}
\definecolor {cornsilk3}           {rgb}{0.80,0.78,0.69}
\definecolor {cornsilk4}           {rgb}{0.55,0.53,0.47}
\definecolor {ivory2}              {rgb}{0.93,0.93,0.88}
\definecolor {ivory3}              {rgb}{0.80,0.80,0.76}
\definecolor {ivory4}              {rgb}{0.55,0.55,0.51}
\definecolor {honeydew2}           {rgb}{0.88,0.93,0.88}
\definecolor {honeydew3}           {rgb}{0.76,0.80,0.76}
\definecolor {honeydew4}           {rgb}{0.51,0.55,0.51}
\definecolor {lavenderblush2}      {rgb}{0.93,0.88,0.90}
\definecolor {lavenderblush3}      {rgb}{0.80,0.76,0.77}
\definecolor {lavenderblush4}      {rgb}{0.55,0.51,0.53}
\definecolor {mistyrose2}          {rgb}{0.93,0.84,0.82}
\definecolor {mistyrose3}          {rgb}{0.80,0.72,0.71}
\definecolor {mistyrose4}          {rgb}{0.55,0.49,0.48}
\definecolor {azure2}              {rgb}{0.88,0.93,0.93}
\definecolor {azure3}              {rgb}{0.76,0.80,0.80}
\definecolor {azure4}              {rgb}{0.51,0.55,0.55}
\definecolor {slateblue1}          {rgb}{0.51,0.44,1.00}
\definecolor {slateblue2}          {rgb}{0.48,0.40,0.93}
\definecolor {slateblue3}          {rgb}{0.41,0.35,0.80}
\definecolor {slateblue4}          {rgb}{0.28,0.24,0.55}
\definecolor {royalblue1}          {rgb}{0.28,0.46,1.00}
\definecolor {royalblue2}          {rgb}{0.26,0.43,0.93}
\definecolor {royalblue3}          {rgb}{0.23,0.37,0.80}
\definecolor {royalblue4}          {rgb}{0.15,0.25,0.55}
\definecolor {blue2}               {rgb}{0.00,0.00,0.93}
\definecolor {blue4}               {rgb}{0.00,0.00,0.55}
\definecolor {dodgerblue2}         {rgb}{0.11,0.53,0.93}
\definecolor {dodgerblue3}         {rgb}{0.09,0.45,0.80}
\definecolor {dodgerblue4}         {rgb}{0.06,0.31,0.55}
\definecolor {steelblue1}          {rgb}{0.39,0.72,1.00}
\definecolor {steelblue2}          {rgb}{0.36,0.67,0.93}
\definecolor {steelblue3}          {rgb}{0.31,0.58,0.80}
\definecolor {steelblue4}          {rgb}{0.21,0.39,0.55}
\definecolor {deepskyblue2}        {rgb}{0.00,0.70,0.93}
\definecolor {deepskyblue3}        {rgb}{0.00,0.60,0.80}
\definecolor {deepskyblue4}        {rgb}{0.00,0.41,0.55}
\definecolor {skyblue1}            {rgb}{0.53,0.81,1.00}
\definecolor {skyblue2}            {rgb}{0.49,0.75,0.93}
\definecolor {skyblue3}            {rgb}{0.42,0.65,0.80}
\definecolor {skyblue4}            {rgb}{0.29,0.44,0.55}
\definecolor {lightskyblue1}       {rgb}{0.69,0.89,1.00}
\definecolor {lightskyblue2}       {rgb}{0.64,0.83,0.93}
\definecolor {lightskyblue3}       {rgb}{0.55,0.71,0.80}
\definecolor {lightskyblue4}       {rgb}{0.38,0.48,0.55}
\definecolor {slategray1}          {rgb}{0.78,0.89,1.00}
\definecolor {slategray2}          {rgb}{0.73,0.83,0.93}
\definecolor {slategray3}          {rgb}{0.62,0.71,0.80}
\definecolor {slategray4}          {rgb}{0.42,0.48,0.55}
\definecolor {lightsteelblue1}     {rgb}{0.79,0.88,1.00}
\definecolor {lightsteelblue2}     {rgb}{0.74,0.82,0.93}
\definecolor {lightsteelblue3}     {rgb}{0.64,0.71,0.80}
\definecolor {lightsteelblue4}     {rgb}{0.43,0.48,0.55}
\definecolor {lightblue1}          {rgb}{0.75,0.94,1.00}
\definecolor {lightblue2}          {rgb}{0.70,0.87,0.93}
\definecolor {lightblue3}          {rgb}{0.60,0.75,0.80}
\definecolor {lightblue4}          {rgb}{0.41,0.51,0.55}
\definecolor {lightcyan2}          {rgb}{0.82,0.93,0.93}
\definecolor {lightcyan3}          {rgb}{0.71,0.80,0.80}
\definecolor {lightcyan4}          {rgb}{0.48,0.55,0.55}
\definecolor {paleturquoise1}      {rgb}{0.73,1.00,1.00}
\definecolor {paleturquoise2}      {rgb}{0.68,0.93,0.93}
\definecolor {paleturquoise3}      {rgb}{0.59,0.80,0.80}
\definecolor {paleturquoise4}      {rgb}{0.40,0.55,0.55}
\definecolor {cadetblue1}          {rgb}{0.60,0.96,1.00}
\definecolor {cadetblue2}          {rgb}{0.56,0.90,0.93}
\definecolor {cadetblue3}          {rgb}{0.48,0.77,0.80}
\definecolor {cadetblue4}          {rgb}{0.33,0.53,0.55}
\definecolor {turquoise1}          {rgb}{0.00,0.96,1.00}
\definecolor {turquoise2}          {rgb}{0.00,0.90,0.93}
\definecolor {turquoise3}          {rgb}{0.00,0.77,0.80}
\definecolor {turquoise4}          {rgb}{0.00,0.53,0.55}
\definecolor {cyan2}               {rgb}{0.00,0.93,0.93}
\definecolor {cyan3}               {rgb}{0.00,0.80,0.80}
\definecolor {cyan4}               {rgb}{0.00,0.55,0.55}
\definecolor {darkslategray1}      {rgb}{0.59,1.00,1.00}
\definecolor {darkslategray2}      {rgb}{0.55,0.93,0.93}
\definecolor {darkslategray3}      {rgb}{0.47,0.80,0.80}
\definecolor {darkslategray4}      {rgb}{0.32,0.55,0.55}
\definecolor {aquamarine2}         {rgb}{0.46,0.93,0.78}
\definecolor {aquamarine4}         {rgb}{0.27,0.55,0.45}
\definecolor {darkseagreen1}       {rgb}{0.76,1.00,0.76}
\definecolor {darkseagreen2}       {rgb}{0.71,0.93,0.71}
\definecolor {darkseagreen3}       {rgb}{0.61,0.80,0.61}
\definecolor {darkseagreen4}       {rgb}{0.41,0.55,0.41}
\definecolor {seagreen1}           {rgb}{0.33,1.00,0.62}
\definecolor {seagreen2}           {rgb}{0.31,0.93,0.58}
\definecolor {seagreen3}           {rgb}{0.26,0.80,0.50}
\definecolor {palegreen1}          {rgb}{0.60,1.00,0.60}
\definecolor {palegreen2}          {rgb}{0.56,0.93,0.56}
\definecolor {palegreen3}          {rgb}{0.49,0.80,0.49}
\definecolor {palegreen4}          {rgb}{0.33,0.55,0.33}
\definecolor {springgreen2}        {rgb}{0.00,0.93,0.46}
\definecolor {springgreen3}        {rgb}{0.00,0.80,0.40}
\definecolor {springgreen4}        {rgb}{0.00,0.55,0.27}
\definecolor {green2}              {rgb}{0.00,0.93,0.00}
\definecolor {green3}              {rgb}{0.00,0.80,0.00}
\definecolor {green4}              {rgb}{0.00,0.55,0.00}
\definecolor {chartreuse2}         {rgb}{0.46,0.93,0.00}
\definecolor {chartreuse3}         {rgb}{0.40,0.80,0.00}
\definecolor {chartreuse4}         {rgb}{0.27,0.55,0.00}
\definecolor {olivedrab1}          {rgb}{0.75,1.00,0.24}
\definecolor {olivedrab2}          {rgb}{0.70,0.93,0.23}
\definecolor {olivedrab4}          {rgb}{0.41,0.55,0.13}
\definecolor {darkolivegreen1}     {rgb}{0.79,1.00,0.44}
\definecolor {darkolivegreen2}     {rgb}{0.74,0.93,0.41}
\definecolor {darkolivegreen3}     {rgb}{0.64,0.80,0.35}
\definecolor {darkolivegreen4}     {rgb}{0.43,0.55,0.24}
\definecolor {khaki1}              {rgb}{1.00,0.96,0.56}
\definecolor {khaki2}              {rgb}{0.93,0.90,0.52}
\definecolor {khaki3}              {rgb}{0.80,0.78,0.45}
\definecolor {khaki4}              {rgb}{0.55,0.53,0.31}
\definecolor {lightgoldenrod1}     {rgb}{1.00,0.93,0.55}
\definecolor {lightgoldenrod2}     {rgb}{0.93,0.86,0.51}
\definecolor {lightgoldenrod3}     {rgb}{0.80,0.75,0.44}
\definecolor {lightgoldenrod4}     {rgb}{0.55,0.51,0.30}
\definecolor {lightyellow2}        {rgb}{0.93,0.93,0.82}
\definecolor {lightyellow3}        {rgb}{0.80,0.80,0.71}
\definecolor {lightyellow4}        {rgb}{0.55,0.55,0.48}
\definecolor {yellow2}             {rgb}{0.93,0.93,0.00}
\definecolor {yellow3}             {rgb}{0.80,0.80,0.00}
\definecolor {yellow4}             {rgb}{0.55,0.55,0.00}
\definecolor {gold2}               {rgb}{0.93,0.79,0.00}
\definecolor {gold3}               {rgb}{0.80,0.68,0.00}
\definecolor {gold4}               {rgb}{0.55,0.46,0.00}
\definecolor {goldenrod1}          {rgb}{1.00,0.76,0.15}
\definecolor {goldenrod2}          {rgb}{0.93,0.71,0.13}
\definecolor {goldenrod3}          {rgb}{0.80,0.61,0.11}
\definecolor {goldenrod4}          {rgb}{0.55,0.41,0.08}
\definecolor {darkgoldenrod1}      {rgb}{1.00,0.73,0.06}
\definecolor {darkgoldenrod2}      {rgb}{0.93,0.68,0.05}
\definecolor {darkgoldenrod3}      {rgb}{0.80,0.58,0.05}
\definecolor {darkgoldenrod4}      {rgb}{0.55,0.40,0.03}
\definecolor {rosybrown1}          {rgb}{1.00,0.76,0.76}
\definecolor {rosybrown2}          {rgb}{0.93,0.71,0.71}
\definecolor {rosybrown3}          {rgb}{0.80,0.61,0.61}
\definecolor {rosybrown4}          {rgb}{0.55,0.41,0.41}
\definecolor {indianred1}          {rgb}{1.00,0.42,0.42}
\definecolor {indianred2}          {rgb}{0.93,0.39,0.39}
\definecolor {indianred3}          {rgb}{0.80,0.33,0.33}
\definecolor {indianred4}          {rgb}{0.55,0.23,0.23}
\definecolor {sienna1}             {rgb}{1.00,0.51,0.28}
\definecolor {sienna2}             {rgb}{0.93,0.47,0.26}
\definecolor {sienna3}             {rgb}{0.80,0.41,0.22}
\definecolor {sienna4}             {rgb}{0.55,0.28,0.15}
\definecolor {burlywood1}          {rgb}{1.00,0.83,0.61}
\definecolor {burlywood2}          {rgb}{0.93,0.77,0.57}
\definecolor {burlywood3}          {rgb}{0.80,0.67,0.49}
\definecolor {burlywood4}          {rgb}{0.55,0.45,0.33}
\definecolor {wheat1}              {rgb}{1.00,0.91,0.73}
\definecolor {wheat2}              {rgb}{0.93,0.85,0.68}
\definecolor {wheat3}              {rgb}{0.80,0.73,0.59}
\definecolor {wheat4}              {rgb}{0.55,0.49,0.40}
\definecolor {tan1}                {rgb}{1.00,0.65,0.31}
\definecolor {tan2}                {rgb}{0.93,0.60,0.29}
\definecolor {tan4}                {rgb}{0.55,0.35,0.17}
\definecolor {chocolate1}          {rgb}{1.00,0.50,0.14}
\definecolor {chocolate2}          {rgb}{0.93,0.46,0.13}
\definecolor {chocolate3}          {rgb}{0.80,0.40,0.11}
\definecolor {firebrick1}          {rgb}{1.00,0.19,0.19}
\definecolor {firebrick2}          {rgb}{0.93,0.17,0.17}
\definecolor {firebrick3}          {rgb}{0.80,0.15,0.15}
\definecolor {firebrick4}          {rgb}{0.55,0.10,0.10}
\definecolor {brown1}              {rgb}{1.00,0.25,0.25}
\definecolor {brown2}              {rgb}{0.93,0.23,0.23}
\definecolor {brown3}              {rgb}{0.80,0.20,0.20}
\definecolor {brown4}              {rgb}{0.55,0.14,0.14}
\definecolor {salmon1}             {rgb}{1.00,0.55,0.41}
\definecolor {salmon2}             {rgb}{0.93,0.51,0.38}
\definecolor {salmon3}             {rgb}{0.80,0.44,0.33}
\definecolor {salmon4}             {rgb}{0.55,0.30,0.22}
\definecolor {lightsalmon2}        {rgb}{0.93,0.58,0.45}
\definecolor {lightsalmon3}        {rgb}{0.80,0.51,0.38}
\definecolor {lightsalmon4}        {rgb}{0.55,0.34,0.26}
\definecolor {orange2}             {rgb}{0.93,0.60,0.00}
\definecolor {orange3}             {rgb}{0.80,0.52,0.00}
\definecolor {orange4}             {rgb}{0.55,0.35,0.00}
\definecolor {darkorange1}         {rgb}{1.00,0.50,0.00}
\definecolor {darkorange2}         {rgb}{0.93,0.46,0.00}
\definecolor {darkorange3}         {rgb}{0.80,0.40,0.00}
\definecolor {darkorange4}         {rgb}{0.55,0.27,0.00}
\definecolor {coral1}              {rgb}{1.00,0.45,0.34}
\definecolor {coral2}              {rgb}{0.93,0.42,0.31}
\definecolor {coral3}              {rgb}{0.80,0.36,0.27}
\definecolor {coral4}              {rgb}{0.55,0.24,0.18}
\definecolor {tomato2}             {rgb}{0.93,0.36,0.26}
\definecolor {tomato3}             {rgb}{0.80,0.31,0.22}
\definecolor {tomato4}             {rgb}{0.55,0.21,0.15}
\definecolor {orangered2}          {rgb}{0.93,0.25,0.00}
\definecolor {orangered3}          {rgb}{0.80,0.22,0.00}
\definecolor {orangered4}          {rgb}{0.55,0.15,0.00}
\definecolor {red2}                {rgb}{0.93,0.00,0.00}
\definecolor {red3}                {rgb}{0.80,0.00,0.00}
\definecolor {red4}                {rgb}{0.55,0.00,0.00}
\definecolor {deeppink2}           {rgb}{0.93,0.07,0.54}
\definecolor {deeppink3}           {rgb}{0.80,0.06,0.46}
\definecolor {deeppink4}           {rgb}{0.55,0.04,0.31}
\definecolor {hotpink1}            {rgb}{1.00,0.43,0.71}
\definecolor {hotpink2}            {rgb}{0.93,0.42,0.65}
\definecolor {hotpink3}            {rgb}{0.80,0.38,0.56}
\definecolor {hotpink4}            {rgb}{0.55,0.23,0.38}
\definecolor {pink1}               {rgb}{1.00,0.71,0.77}
\definecolor {pink2}               {rgb}{0.93,0.66,0.72}
\definecolor {pink3}               {rgb}{0.80,0.57,0.62}
\definecolor {pink4}               {rgb}{0.55,0.39,0.42}
\definecolor {lightpink1}          {rgb}{1.00,0.68,0.73}
\definecolor {lightpink2}          {rgb}{0.93,0.64,0.68}
\definecolor {lightpink3}          {rgb}{0.80,0.55,0.58}
\definecolor {lightpink4}          {rgb}{0.55,0.37,0.40}
\definecolor {palevioletred1}      {rgb}{1.00,0.51,0.67}
\definecolor {palevioletred2}      {rgb}{0.93,0.47,0.62}
\definecolor {palevioletred3}      {rgb}{0.80,0.41,0.54}
\definecolor {palevioletred4}      {rgb}{0.55,0.28,0.36}
\definecolor {maroon1}             {rgb}{1.00,0.20,0.70}
\definecolor {maroon2}             {rgb}{0.93,0.19,0.65}
\definecolor {maroon3}             {rgb}{0.80,0.16,0.56}
\definecolor {maroon4}             {rgb}{0.55,0.11,0.38}
\definecolor {violetred1}          {rgb}{1.00,0.24,0.59}
\definecolor {violetred2}          {rgb}{0.93,0.23,0.55}
\definecolor {violetred3}          {rgb}{0.80,0.20,0.47}
\definecolor {violetred4}          {rgb}{0.55,0.13,0.32}
\definecolor {magenta2}            {rgb}{0.93,0.00,0.93}
\definecolor {magenta3}            {rgb}{0.80,0.00,0.80}
\definecolor {magenta4}            {rgb}{0.55,0.00,0.55}
\definecolor {orchid1}             {rgb}{1.00,0.51,0.98}
\definecolor {orchid2}             {rgb}{0.93,0.48,0.91}
\definecolor {orchid3}             {rgb}{0.80,0.41,0.79}
\definecolor {orchid4}             {rgb}{0.55,0.28,0.54}
\definecolor {plum1}               {rgb}{1.00,0.73,1.00}
\definecolor {plum2}               {rgb}{0.93,0.68,0.93}
\definecolor {plum3}               {rgb}{0.80,0.59,0.80}
\definecolor {plum4}               {rgb}{0.55,0.40,0.55}
\definecolor {mediumorchid1}       {rgb}{0.88,0.40,1.00}
\definecolor {mediumorchid2}       {rgb}{0.82,0.37,0.93}
\definecolor {mediumorchid3}       {rgb}{0.71,0.32,0.80}
\definecolor {mediumorchid4}       {rgb}{0.48,0.22,0.55}
\definecolor {darkorchid1}         {rgb}{0.75,0.24,1.00}
\definecolor {darkorchid2}         {rgb}{0.70,0.23,0.93}
\definecolor {darkorchid3}         {rgb}{0.60,0.20,0.80}
\definecolor {darkorchid4}         {rgb}{0.41,0.13,0.55}
\definecolor {purple1}             {rgb}{0.61,0.19,1.00}
\definecolor {purple2}             {rgb}{0.57,0.17,0.93}
\definecolor {purple3}             {rgb}{0.49,0.15,0.80}
\definecolor {purple4}             {rgb}{0.33,0.10,0.55}
\definecolor {mediumpurple1}       {rgb}{0.67,0.51,1.00}
\definecolor {mediumpurple2}       {rgb}{0.62,0.47,0.93}
\definecolor {mediumpurple3}       {rgb}{0.54,0.41,0.80}
\definecolor {mediumpurple4}       {rgb}{0.36,0.28,0.55}
\definecolor {thistle1}            {rgb}{1.00,0.88,1.00}
\definecolor {thistle2}            {rgb}{0.93,0.82,0.93}
\definecolor {thistle3}            {rgb}{0.80,0.71,0.80}
\definecolor {thistle4}            {rgb}{0.55,0.48,0.55}
\definecolor {gray1}               {rgb}{0.01,0.01,0.01}
\definecolor {gray2}               {rgb}{0.02,0.02,0.02}
\definecolor {gray3}               {rgb}{0.03,0.03,0.03}
\definecolor {gray4}               {rgb}{0.04,0.04,0.04}
\definecolor {gray5}               {rgb}{0.05,0.05,0.05}
\definecolor {gray6}               {rgb}{0.06,0.06,0.06}
\definecolor {gray7}               {rgb}{0.07,0.07,0.07}
\definecolor {gray8}               {rgb}{0.08,0.08,0.08}
\definecolor {gray9}               {rgb}{0.09,0.09,0.09}
\definecolor {gray10}              {rgb}{0.10,0.10,0.10}
\definecolor {gray11}              {rgb}{0.11,0.11,0.11}
\definecolor {gray12}              {rgb}{0.12,0.12,0.12}
\definecolor {gray13}              {rgb}{0.13,0.13,0.13}
\definecolor {gray14}              {rgb}{0.14,0.14,0.14}
\definecolor {gray15}              {rgb}{0.15,0.15,0.15}
\definecolor {gray16}              {rgb}{0.16,0.16,0.16}
\definecolor {gray17}              {rgb}{0.17,0.17,0.17}
\definecolor {gray18}              {rgb}{0.18,0.18,0.18}
\definecolor {gray19}              {rgb}{0.19,0.19,0.19}
\definecolor {gray20}              {rgb}{0.20,0.20,0.20}
\definecolor {gray21}              {rgb}{0.21,0.21,0.21}
\definecolor {gray22}              {rgb}{0.22,0.22,0.22}
\definecolor {gray23}              {rgb}{0.23,0.23,0.23}
\definecolor {gray24}              {rgb}{0.24,0.24,0.24}
\definecolor {gray25}              {rgb}{0.25,0.25,0.25}
\definecolor {gray26}              {rgb}{0.26,0.26,0.26}
\definecolor {gray27}              {rgb}{0.27,0.27,0.27}
\definecolor {gray28}              {rgb}{0.28,0.28,0.28}
\definecolor {gray29}              {rgb}{0.29,0.29,0.29}
\definecolor {gray30}              {rgb}{0.30,0.30,0.30}
\definecolor {gray31}              {rgb}{0.31,0.31,0.31}
\definecolor {gray32}              {rgb}{0.32,0.32,0.32}
\definecolor {gray33}              {rgb}{0.33,0.33,0.33}
\definecolor {gray34}              {rgb}{0.34,0.34,0.34}
\definecolor {gray35}              {rgb}{0.35,0.35,0.35}
\definecolor {gray36}              {rgb}{0.36,0.36,0.36}
\definecolor {gray37}              {rgb}{0.37,0.37,0.37}
\definecolor {gray38}              {rgb}{0.38,0.38,0.38}
\definecolor {gray39}              {rgb}{0.39,0.39,0.39}
\definecolor {gray40}              {rgb}{0.40,0.40,0.40}
\definecolor {gray42}              {rgb}{0.42,0.42,0.42}
\definecolor {gray43}              {rgb}{0.43,0.43,0.43}
\definecolor {gray44}              {rgb}{0.44,0.44,0.44}
\definecolor {gray45}              {rgb}{0.45,0.45,0.45}
\definecolor {gray46}              {rgb}{0.46,0.46,0.46}
\definecolor {gray47}              {rgb}{0.47,0.47,0.47}
\definecolor {gray48}              {rgb}{0.48,0.48,0.48}
\definecolor {gray49}              {rgb}{0.49,0.49,0.49}
\definecolor {gray50}              {rgb}{0.50,0.50,0.50}
\definecolor {gray51}              {rgb}{0.51,0.51,0.51}
\definecolor {gray52}              {rgb}{0.52,0.52,0.52}
\definecolor {gray53}              {rgb}{0.53,0.53,0.53}
\definecolor {gray54}              {rgb}{0.54,0.54,0.54}
\definecolor {gray55}              {rgb}{0.55,0.55,0.55}
\definecolor {gray56}              {rgb}{0.56,0.56,0.56}
\definecolor {gray57}              {rgb}{0.57,0.57,0.57}
\definecolor {gray58}              {rgb}{0.58,0.58,0.58}
\definecolor {gray59}              {rgb}{0.59,0.59,0.59}
\definecolor {gray60}              {rgb}{0.60,0.60,0.60}
\definecolor {gray61}              {rgb}{0.61,0.61,0.61}
\definecolor {gray62}              {rgb}{0.62,0.62,0.62}
\definecolor {gray63}              {rgb}{0.63,0.63,0.63}
\definecolor {gray64}              {rgb}{0.64,0.64,0.64}
\definecolor {gray65}              {rgb}{0.65,0.65,0.65}
\definecolor {gray66}              {rgb}{0.66,0.66,0.66}
\definecolor {gray67}              {rgb}{0.67,0.67,0.67}
\definecolor {gray68}              {rgb}{0.68,0.68,0.68}
\definecolor {gray69}              {rgb}{0.69,0.69,0.69}
\definecolor {gray70}              {rgb}{0.70,0.70,0.70}
\definecolor {gray71}              {rgb}{0.71,0.71,0.71}
\definecolor {gray72}              {rgb}{0.72,0.72,0.72}
\definecolor {gray73}              {rgb}{0.73,0.73,0.73}
\definecolor {gray74}              {rgb}{0.74,0.74,0.74}
\definecolor {gray75}              {rgb}{0.75,0.75,0.75}
\definecolor {gray76}              {rgb}{0.76,0.76,0.76}
\definecolor {gray77}              {rgb}{0.77,0.77,0.77}
\definecolor {gray78}              {rgb}{0.78,0.78,0.78}
\definecolor {gray79}              {rgb}{0.79,0.79,0.79}
\definecolor {gray80}              {rgb}{0.80,0.80,0.80}
\definecolor {gray81}              {rgb}{0.81,0.81,0.81}
\definecolor {gray82}              {rgb}{0.82,0.82,0.82}
\definecolor {gray83}              {rgb}{0.83,0.83,0.83}
\definecolor {gray84}              {rgb}{0.84,0.84,0.84}
\definecolor {gray85}              {rgb}{0.85,0.85,0.85}
\definecolor {gray86}              {rgb}{0.86,0.86,0.86}
\definecolor {gray87}              {rgb}{0.87,0.87,0.87}
\definecolor {gray88}              {rgb}{0.88,0.88,0.88}
\definecolor {gray89}              {rgb}{0.89,0.89,0.89}
\definecolor {gray90}              {rgb}{0.90,0.90,0.90}
\definecolor {gray91}              {rgb}{0.91,0.91,0.91}
\definecolor {gray92}              {rgb}{0.92,0.92,0.92}
\definecolor {gray93}              {rgb}{0.93,0.93,0.93}
\definecolor {gray94}              {rgb}{0.94,0.94,0.94}
\definecolor {gray95}              {rgb}{0.95,0.95,0.95}
\definecolor {gray97}              {rgb}{0.97,0.97,0.97}
\definecolor {gray98}              {rgb}{0.98,0.98,0.98}
\definecolor {gray99}              {rgb}{0.99,0.99,0.99}
\definecolor {darkgrey}            {rgb}{0.66,0.66,0.66}

%% file: macros/rs_macros_optsmt.tex
\newcommand{\omt}{\ensuremath{\text{OMT}}\xspace}

\newcommand{\cost}{\ensuremath{{cost}}\xspace}

\newcommand\mysout{\bgroup \markoverwith{{-}}\ULon}
\newcommand\nosout{\bgroup \markoverwith{{ }}\ULon}
\definecolor{mygray}{rgb}{0.90,0.90,0.90}
\definecolor{mywhite}{rgb}{1.00,1.00,1.00}

\newcommand{\optimathsat}{\textsc{OptiMathSAT}\xspace}

\newcommand{\symba}{\textsc{Symba}\xspace}

\newcommand{\bclt}{\textsc{bclt}\xspace}

%% file: macros/rs_macros_sat.tex
\newcommand{\satres}{\textsc{sat}\xspace}
\newcommand{\unsatres}{\textsc{unsat}\xspace}


%% file: macros/rs_macros_smt.tex
\newcommand{\vi}{\ensuremath{\varphi}\xspace}

\newcommand{\cp}{\ensuremath{c^p}\xspace}

\newcommand{\smt}{SMT\xspace}

\newcommand{\fl}{\ensuremath{\mathcal{FP}}\xspace}

\newcommand{\nlarat}{\ensuremath{\mathcal{NLRA}}\xspace}
\newcommand{\nlaint}{\ensuremath{\mathcal{NLIA}}\xspace}
\newcommand{\bv}{\ensuremath{\mathcal{BV}}\xspace}
\newcommand{\mem}{\ensuremath{\mathcal{AR}}\xspace}

\newcommand{\zthree}{\textsc{Z3}\xspace}

\newcommand{\mathsatfive}{\textsc{MathSAT5}\xspace}


%% file: macros/rs_macros_long.tex
\ifthenelse{\equal{\longversion}{true}}
{%
  \renewcommand{\ignoreinshort}[1]{{\textcolor{black}{#1}}}
  \renewcommand{\ignoreinlong}[1]{}
  \specialcomment{IGNOREINSHORT}{\begingroup\color{black}}{\endgroup}
  \excludecomment{IGNOREINLONG}
}%
{%
 \renewcommand{\ignoreinshort}[1]{}
 \renewcommand{\ignoreinlong}[1]{#1}
 \excludecomment{IGNOREINSHORT}
  \specialcomment{IGNOREINLONG}{\begingroup\color{black}}{\endgroup}
}
 \excludecomment{IGNORE}

%% file: macros/pt_macros_abrv.tex
\newcommand{\smtlib}{\textsc{SMT-LIB}}
\newcommand{\flatzinc}{\textsc{FlatZinc}}
\newcommand{\minizinc}{\textsc{MiniZinc}}

\newcommand{\mznTofzn}{\textsc{mzn2fzn}}

\newcommand{\fzntoomt}{\textsc{fzn2omt}}
\newcommand{\mzntofzn}{\textsc{mzn2fzn}}
\newcommand{\emzntofzn}{\textsc{emzn2fzn}}
\newcommand{\fzntosmt}{\textsc{fzn2smt}}
\newcommand{\omttomzn}{\textsc{omt2mzn}}
\newcommand{\minisearch}{\textsc{MiniSearch}}

\newcommand{\hazel}{\textsc{Hazel}}
\newcommand{\puli}{\textsc{Puli}}

\newcommand{\cegio}{\textsc{Cegio}}

\newcommand{\choco}{\textsc{Choco}}
\newcommand{\chuffed}{\textsc{Chuffed}}
\newcommand{\gfd}{\textsc{G12(fd)}}
\newcommand{\gmip}{\textsc{G12(mip)}}
\newcommand{\gecode}{\textsc{Gecode}}
\newcommand{\gurobi}{\textsc{Gurobi}}
\newcommand{\hcsp}{\textsc{HaifaCSP}}
\newcommand{\izplus}{\textsc{iZplus}}
\newcommand{\jacop}{\textsc{JaCoP}}

\newcommand{\ortools}{\textsc{OR-Tools}}
\newcommand{\ortoolssat}{\textsc{OR-Tools(sat)}}

\newcommand{\picat}{\textsc{Picat}}
\newcommand{\picatsat}{\textsc{Picat(sat)}}
\newcommand{\picatcp}{\textsc{Picat(cp)}}
\newcommand{\vbest}{\textsc{virtual best}}

\renewcommand{\cost}{\ensuremath{{obj}}\xspace}

\newcommand{\fp}{\ensuremath{FP}\xspace}

\newcommand{\sat}{SAT}
\newcommand{\csp}{\textsc{CSP}}

\newcommand{\cop}{\textsc{COP}}
\renewcommand{\cp}{\textsc{CP}}

\newcommand{\maxsat}{\textsc{MaxSAT}}
\newcommand{\maxsmt}{\textsc{MaxSMT}}

\newcommand{\pb}  {\ensuremath{\mathcal{PB}}\xspace}
\newcommand{\lra} {\ensuremath{\mathcal{LRA}}\xspace}
\newcommand{\lia} {\ensuremath{\mathcal{LIA}}\xspace}
\newcommand{\lira}{\ensuremath{\mathcal{LIRA}}\xspace}
\renewcommand{\bv}  {\ensuremath{\mathcal{BV}}\xspace}
\renewcommand{\fp}  {\ensuremath{\mathcal{FP}}\xspace}

\newcommand{\omtpb}{\ensuremath{\text{OMT}(\pb)}\xspace}
\newcommand{\omtlra} {\ensuremath{\text{OMT}(\lra)}\xspace}
\newcommand{\omtlia} {\ensuremath{\text{OMT}(\lia)}\xspace}

\newcommand{\omtbv}  {\ensuremath{\text{OMT}(\bv)}\xspace}

\newcommand{\attreq}[1]{\ensuremath{\ifthenelse{\equal{#1}{}}{A}{A[#1]}}\xspace}
\newcommand{\dattreq}[2]{\ensuremath{\ifthenelse{\equal{#2}{}}{A_{#1}}{A_{#1}[#2]}}\xspace}

\newcommand{\dcala}[1]{\ensuremath{\ifthenelse{\equal{#1}{}}{\cala_{\varphi}}{\cala_{\varphi}[#1]}}\xspace}

\newcommand{\scsat}{\textbf{s}}
\newcommand{\scopt}{\textbf{o}}

\renewcommand{\cost}{{\sf obj}\xspace}


%% file: macros/pt_macros_notes.tex
\newenvironment{ptchange}{\color{darkviolet}}{\normalcolor}

%% file: macros/pt_macros_lst.tex
\lstdefinestyle{box}{
    backgroundcolor=\transparent{0.25}\color{lightgray},
    framexleftmargin=0pt,
    framexrightmargin=0pt,
    framextopmargin=0pt,
    framexbottommargin=0pt,
    frame=tblr,
    framerule=0.25pt,
    basicstyle=\small\ttfamily
}

\lstdefinestyle{bigbox}{
    backgroundcolor=\transparent{0.25}\color{lightgray},
    framexleftmargin=0pt,
    framexrightmargin=0pt,
    framextopmargin=0pt,
    framexbottommargin=0pt,
    frame=tblr,
    framerule=0.25pt,
    basicstyle=\footnotesize\ttfamily,
    numbers=left,
    stepnumber=1,
    numberstyle=\footnotesize\ttfamily,
    numbersep=10pt,
    showstringspaces=false,
    tabsize=1,
    breaklines=true,
    breakatwhitespace=false,
    xleftmargin=3em,
    framexleftmargin=2.5em
}

\lstdefinestyle{numbers}{
    numbers=left,
    stepnumber=1,
    numberstyle=\footnotesize\ttfamily,
    numbersep=10pt,
    showstringspaces=false,
    tabsize=1,
    breaklines=true,
    breakatwhitespace=false,
    xleftmargin=3em,
    framexleftmargin=2.5em
}

\lstdefinestyle{escape}{
    escapeinside={(*}{*)}
}

\lstnewenvironment{pseudocode}{\lstset{language=C++,style=box,style=numbers,style=escape}} {}


%% file: macros/pt_macros_cleveref.tex
\crefformat{footnote}{#2\footnotemark[#1]#3}

%% file: macros/pt_macros_alg.tex
\renewcommand\algorithmiccomment[1]{%
  \hfill\#\ \eqparbox{COMMENT}{#1}%
}

\makeatletter
\patchcmd{\algorithmic}{\addtolength{\ALC@tlm}{\leftmargin} }{\addtolength{\ALC@tlm}{\leftmargin}}{}{}
\makeatother

\SetCommentSty{mycommfont}

%% file: macros/pt_macros_proofs.tex
\ifthenelse{\equal{\withproofs}{true}}
{%
\specialcomment{PROOF}{\begingroup\color{darkviolet}}{\endgroup}
\excludecomment{APPENDIX}
}%
{%
\specialcomment{APPENDIX}{\begingroup\color{black}}{\endgroup}
\excludecomment{PROOF}
}

%% file: macros/macros_anon.tex
\newcommand{\realoptimathsat}{\textsc{OptiMathSAT}\xspace}

%% file: macros/fc_macros_notes.tex
\newcommand{\pysmt}{\textsc{pySMT}}

\newcommand{\optimsmt}{\textsc{\optimathsat{}(smt)}}
\newcommand{\optimfzn}{\textsc{\optimathsat{}(fzn)}}
\newcommand{\optimfzne}{\textsc{\optimathsat{}(fzn+e)}}
\newcommand{\vbestpar}[1]{\textsc{Virtual Best(#1)}}

\definecolor{inchworm}{rgb}{0.7, 0.93, 0.36}
\definecolor{khaki)}{rgb}{0.94, 0.9, 0.55}

\newcolumntype{x}{>{\columncolor{inchworm}}r}
\newcolumntype{y}{>{\columncolor{khaki}}r}

\newcolumntype{R}{>{\columncolor{inchworm}}r}
\newcolumntype{C}{>{\columncolor{inchworm}}c}
\newcolumntype{L}{>{\columncolor{inchworm}}l}

%% file: src/abstract.tex
Optimization Modulo Theories (OMT) is an extension of SMT that allows
for finding models that optimize objective functions.
In this paper we aim at bridging the gap between Constraint
Programming (CP) and OMT, {\em in
both directions}.
%
First, we have extended the 
OMT
solver \optimathsat{} with a \flatzinc{} interface --
which can also be used as \flatzinc{}-to-OMT encoder for
other OMT solvers. 
This allows OMT tools to be used in combination with \mznTofzn{}
on the large amount of CP problems coming from the \minizinc{}
community.
Second, we have introduced a tool for translating SMT and OMT problems
on the linear arithmetic and bit-vector theories into \minizinc.  This
allows \minizinc{} solvers to be used on a large amount of SMT/OMT
problems. 

We have discussed the main issues we had to cope with in either
directions.  We have performed an extensive empirical evaluation
comparing three state-of-the-art OMT-based tools
with many state-of-the-art CP tools on (i) CP
problems coming from the \minizinc{} challenge, and
(ii) OMT problems coming mostly from formal verification. 
This analysis also allowed us to identify 
some criticalities, in terms of efficiency and correctness,
one has to cope with when addressing CP problems with OMT tools, and
vice versa. 


\ignore{This analysis allowed us to identify 
some criticalities, in terms of efficiency and correctness,
 either community has to cope with when addressing problems coming from the
 other community.}

%% file: src/introduction.tex

The last two decades have witnessed the rise of Satisfiability Modulo Theories
(\smt{}) \cite{BSST09HBSAT} as efficient tool
for dealing with several applications of industrial interest, in
particular in the contexts
of Formal Verification (FV). 
\smt{} is the problem of finding 
value assignments satisfying some formula in first-order logic 
wrt. some background theory.
Optimization Modulo Theories (\omt{})
\cite{nieuwenhuis_sat06,st-ijcar12,%
li_popl14,LarrazORR14,
st_tacas15,bjorner_tacas15,NadelR16,%
kovas18}
is a more-recent extension of \smt{}
searching for the \emph{optimal} value assignment(s) w.r.t.  some objective function(s), by means of a combination of \smt{} and optimization
procedures.
(Since \omt{} extends  \smt{}, \ignoreinshort{and that all \omt{}
tools are built on top of \smt{} solvers,} hereafter we often simply say
``\omt{}'' for both \smt{} and \omt{}.)

Several distinctive traits of \omt{} solvers --like, e.g.,
the efficient combination of Boolean and arithmetical reasoning,
incrementality,
the availability of
decision procedures for infinite-precision arithmetic and the ability
to produce conflict explanations-- are a direct consequence of their
tight relationship with the FV domain and its practical needs.
%
On the whole, it appears that \omt{} can be a potentially interesting and
efficient technology for dealing with Constraint Programming (\cp{})
problems as well.
At the same time, modeling \cp{} problems for \omt{} solvers requires a
higher-level of expertise, because the same \cp{} instance can have many
possible alternative formulations, s.t. the performance of \smt{} solvers
on each encoding are hardly predictable \cite{frisch09,frisch10,elgabou14}.

On the other hand, the availability, efficiency and expressiveness of
CP tools makes them of potential interest as backend engines
also for FV applications (e.g.,
\cite{Collavizza2006,Collavizza2010,Grinchtein2015}), in particular
with SW verification, where currently SMT is the dominating backend
technology, s.t. a large amount of SMT-encoded FV problems are
available \cite{brst10_smtlib}.  

\ignore{
\RSNOTE{Hint: 1) provides community Y for a novel spectrum of
technologies from community X to solve its problems, 
2) provides community X with a large interesting
previously-inaccessible problems from community Y}

\RSNOTE{Aggiungere motivazione anche per OMT-TO-CP.
}

\RSNOTE{Hint: 2
components: Boolean (combinatorial) and arithmetical}

\RSTODO{Link/Reference to extended version of the paper}
}

\smallskip
In this paper we aim at bridging the gap between \cp{} and \omt{}, {\em in
both directions}.

In the \cp{}-to-\omt{} direction, we have extended the state-of-the-art \omt{} solver
\optimathsat{} \cite{st_jar18}
with a \flatzinc{} interface (namely ``\fzntoomt{}'').
In combination with the standard \mznTofzn{} encoder \cite{minizinc_url},
this new interface can be used to either (i) solve \cp{} models with
\optimathsat{} directly or (ii) generate \omt{} formulas encoded in
the \smtlib{} \cite{cts_cpaior20_extended} format with optimization
extensions, to be fed to other \omt{} solvers,
such as
\bclt{} \cite{Bofilletal2008CAV}
and
\zthree \cite{bjorner_tacas15}.
%
This allows state-of-the-art \omt{} technology to
be used on \minizinc{}
problems coming from the  \cp{} community.

In the \omt{}-to-\cp{} direction, we have introduced a tool for translating \smt{}
and \omt{} problems
on the theories of linear arithmetic over the integers and rational
(\lira) and bit-vector (\bv) into \minizinc{}
models (hereafter ``\omttomzn{}'').  This
allows \minizinc{} solvers to be used on \omt{}
problems, giving them access to a large amount of  \omt{}
problems, mostly coming from formal verification.

With both directions, we first present and discuss the challenges we
encountered and the solutions we adopted to address the differences
between the two formalisms.
Then we present an extensive empirical evaluation
comparing three \omt{} tools with many state-of-the-art \cp{} tools
on (i) \cp{} problems coming from the \minizinc{} challenge, and
(ii) \omt{} problems coming mostly from formal verification.
This analysis allowed us to identify
some criticalities, in terms of efficiency and correctness,
one has to cope with when addressing \cp{} problems with \omt{} tools, and
vice versa.

Overall,
our new compilers \fzntoomt{} and \omttomzn{} in
combination with the standard compiler \mzntofzn{} \cite{minizinc_url}
provide a framework for
translating problems encoded in the \smtlib{} or the \minizinc{} format
in either direction.
This framework enables also for a comparison between \omt{} solvers and
\cp{} tools on problems that do not belong to their original application
domain.
To the best of our knowledge,
this is the first time that such a framework
has been proposed, and that the \omt{}
and \cp{} technologies have been extensively compared on
problems coming from both fields.


\paragraph{Related Work.}
The tight connection between \smt{} and Constraint Programming (\cp{})
has been known for a relatively long period of time \cite{nieuwenhuisetal2007rta}
and it has previously been subject to investigation. 
Some 
works considered a direct encoding of  \cp{} \cite{frisch09,frisch10}
and weighted \cp{} \cite{ansotegui13} into \smt{} and \maxsmt{}, or an
automatic framework for translating \minizinc{}
--a standard \cp{} modeling language \cite{stuckey_mzn_07}-- into \smtlib{}
--the standard \smt{} format-- \cite{bof10sys,bof12}.
Other works explored the integration of typical \sat{} and \smt{} techniques
within \cp{} solvers \cite{ohriSC09,feydy09}. 
Nowadays, several \minizinc{} solvers
--like, e.g.,  \hcsp{} \cite{vekslerS16} and \picat{} \cite{zhou17}--
are at least partially based on \sat{} technology.

To this extent, our first contribution \fzntoomt{} 
also obviates the loss, due to obsolescence, of the \fzntosmt{}
compiler proposed by Bofill et al. in \cite{bof10sys,bof12}.
%
%
%
%
\fzntosmt{} is not compatible with the changes that have
been introduced to the \minizinc{} and \flatzinc{} standards
starting from version $2.0$ of the \minizinc{} distribution.
Since some of these changes are not backward compatible,
it is also not possible to use \fzntosmt{} in conjunction
with an older version of the \mzntofzn{} compiler when dealing
with recent \minizinc{} models.
Furthermore, \fzntosmt{} translates satisfaction problems into
the Version $1$ of the \smtlib{} standard and produces no \smtlib{}
output in the case of optimization problems, that are solved
directly.
However, the optimization interface of modern \omt{} solvers is
based on the Version $2$ of the \smtlib{} standard.
This makes it difficult to use it together with \omt{} solvers.
Unfortunately, the \fzntosmt{} compiler is closed source, with only the
binaries being freely distributed, and seemingly no longer maintained.
This made it necessary to provide a new alternative solution to \fzntosmt{}.
To this extent, our new \flatzinc{} interface of
\optimathsat{}, \fzntoomt{},
translates both satisfaction and optimization problems
in the Version $2$ of the \smtlib{} standard enriched with the optimization
extensions for \omt{} described in \cite{st_jar18}.


\paragraph{Content.}
The rest of the paper is organized as follows.
In \sref{sec:background} we provide some background on
\omt{}, \minizinc{} and \flatzinc{}.
In \sref{sec:mzn2omt} we describe the process from \minizinc{}
to \omt{}.
In \sref{sec:omt2mzn} we describe the process from \omt{} to \minizinc{}.
In \sref{sec:expeval} we describe an empirical evaluation comparing
a \omt{}-based tool with many state-of-the-art \cp{} tools.
Finally, in \sref{sec:conclusions} we conclude and point out
some further research directions.

A longer and more detailed version of this paper is publicly available
as \cite{cts_cpaior20_extended}.

%% file: src/background_omt.tex

Satisfiability Modulo Theories (\smt{})
is the problem of deciding the satisfiability of a first-order
formula \vi{} with respect to a combination of decidable first-order theories.
Typical theories of \smt{} interest are (the theory of)
 linear arithmetic over the  rationals (\lra{}), the integers (\lia{}) or
their combination (\lira{}), non-linear arithmetic over the rationals (\nlarat{})
or the integers (\nlaint{}), arrays (\mem{}), bit-vectors (\bv{}),
floating-point arithmetic (\fl{}), and their combinations thereof.
(See \cite{nieot-jacm-06,sebastiani07,BSST09HBSAT} for an overview.).
%
The last two decades have witnessed the development of very efficient
\smt{} solvers based on the so-called lazy-\smt{} schema \cite{sebastiani07,BSST09HBSAT}.
This has brought previously-intractable problems to the reach of
state-of-the-art \smt{} solvers.


Optimization Modulo Theories (\omt{}), 
\cite{nieuwenhuis_sat06,st-ijcar12,%
li_popl14,LarrazORR14,st_tacas15,bjorner_tacas15,%
NadelR16,ts_cade19}, 
is an extension to \smt{} that allows for finding a model
of a first-order formula \vi{} that is \emph{optimal} with respect to some objective
function expressed in some background theory, by means of a
combination of \smt{} and optimization procedures. 
State-of-the art OMT tools allow optimization in a variety
of theories, including linear arithmetic over the rationals (\omtlra) \cite{st-ijcar12} and the
integers  (\omtlia) \cite{bjorner_tacas15,st_tacas15}, bit-vectors
(\omtbv) \cite{NadelR16} and floating-point numbers (\omt(\fl{})) \cite{ts_cade19}.

  A relevant strict subcase of \omtlra{}
  is \omt{} with Pseudo-Boolean  objective 
  functions (\omtpb{}) in the form $\sum_i w_i A_i$ s.t.
  $w_i$ are rational values and $A_i$ are Boolean variables whose
  values are interpreted as \set{0,1}.
  Notice that \omtpb{} is also equivalent to (partial weighted)
  \maxsmt{}, the SMT extension of 
  \maxsat{}, and that \omtpb{} and \maxsmt{} can be encoded into \omtlra{}
  but not vice versa \cite{st_tocl14}.
  Encoding \omtpb/\maxsmt{} into \omtlra, however, is not the most
  efficient way to solve them, so that modern \omt{} solvers such as
  \bclt{} \cite{Bofilletal2008CAV}, \optimathsat{} \cite{st_jar18} and
  \zthree{} \cite{bjorner_tacas15} implement  specialized \omtpb/\maxsmt{}
  procedures which are much more efficient than general-purpose
  \omtlra{} ones \cite{bjorner_tacas15,st_tacas17,st_jar18}.

  We stress the fact that ---unlike with purely-combinatorial 
  problems, which are encoded into SAT or \maxsat{} 
  and are thus solved by purely-Boolean search--
  typically \omt{} problems involve the interleaving of
  {\em both} Boolean and arithmetical search: search not
  only for the best truth-value assignment to the
  atomic subformulae, but also
  for the best values to the numerical variables compatible
  with such truth-value assignment \cite{st_tocl14}. 

  \ignore{%
  We stress the fact that ---unlike with purely-combinatorial 
  problems which are encoded into SAT, \maxsat{} or \pb{}-optimization
  and are thus solved by purely-Boolean search--
  typically \omt{} problems involve the interleaving of
  {\em both} Boolean and arithmetical search: not
  only for the best truth-value assignment to the
  atomic subformulae, but also
  for the best values to the numerical variables compatible
  with the truth-value assignment. }


To this date, few \omt{} solvers exist, namely
\bclt{} \cite{Bofilletal2008CAV}, 
\cegio{} \cite{araujo16}, 
\hazel{} \cite{NadelR16},
\realoptimathsat{} 
\cite{st_jar18}, 
\puli{} \cite{kovas18},
\symba{} \cite{li_popl14} and
\zthree{} \cite{bjorner_tacas15}. 
%
To this aim, we observe that
(i) some of these solvers are quite recent,
(ii) most of these solvers focus on different, partially overlapping,
     niche subsets of Optimization Modulo Theories, and 
(iii) the lack of an official Input/Output interface for \omt{}
      makes it hard to compare some of these tools with one another.
\omt finds applications in the context of
static analysis \cite{candeago16,karpenkov17},
formal verification and model checking \cite{liutbt17,ratschan17},
scheduling and planning with resources \cite{kovas18,leofante18,roselli18,oliver18},
software security and requirements engineering \cite{nguyensgm16},
workflow analysis \cite{bertolissi18},
machine learning \cite{teso17}, and
quantum computing \cite{bian17_frocos17}.


\ignoreinlong{A distinctive trait of \smt{} (and \omt{}) solvers is the trade-off of speed against
the ability to certify the correctness of the result of any computation, which is
particularly important in the contexts of Formal Verification (FV) and
Model Checking (MC).
When dealing with linear arithmetic in particular, \smt{} solvers employ
\emph{infinite-precision arithmetic} software libraries to avoid numerical
errors and overflows.
}

\ignoreinshort{\begin{remark}
A distinctive trait of \smt{} (and \omt{}) solvers is the trade-off of speed against
the ability to certify the correctness of the result of any computation, which is
particularly important in the contexts of Formal Verification (FV) and
Model Checking (MC).
When dealing with linear arithmetic in particular, \smt{} solvers employ
\emph{infinite-precision arithmetic} software libraries to avoid numerical
errors and overflows.
\end{remark}}


\smtlib{} \cite{cts_cpaior20_extended} is the standard input format
by \smt{} solvers, it provides a standardized definition of the most prominent theories
supported by \smt{} solvers and the corresponding language primitives to use these
features.
At present, there is no standard input format for modeling optimization problems
targeting \omt{} solvers, although there exist only minor syntactical differences
between the major \omt{} solvers.
The tools presented in this paper conform to the \emph{extended \smtlib{}}
format for \omt{} presented in \cite{st_jar18}, that includes language
primitives for modeling objectives.


\optimathsat{} \cite{st-ijcar12,st_tocl14,st_cav15,st_tacas15,st_tacas17,st_jar18}
is a state-of-the-art \omt{} solver based
on the \mathsatfive{} \smt{} solver \cite{mathsat5-url,mathsat5_tacas13}.
\optimathsat{} features both single- and multi-objective optimization over
arbitrary sets of \lra{}, \lia{}, \lira{}, \bv{}, \fp{}, Pseudo-Boolean (\pb{})
and \maxsmt{} cost functions.
Multiple objective functions can be combined with one another into a Lexicographic
or a Pareto optimization problem, or independently solved in a single run (for the
best efficiency).


%% file: src/background_mzn.tex

\minizinc{} \cite{stuckey_mzn_07,minizinc_url} is a widely adopted high-level
declarative language for modeling Constraint Satisfaction Problems (\csp{})
and Constraint Optimization Problems (\cop{}).
The \minizinc{} format defines three scalar types (bool, int and float)
and two compound types (sets and fixed-size arrays of some scalar type).
The standard provides an extensive list of predefined \emph{global constraints},
a class of high-level language primitives that allows one to encode complex
constraints in a compact way.
\ignoreinshort{%
Furthermore, \minizinc{} supports useful language constructs such as
{\tt if-then-else}, {\tt let} expressions, universal and existential
comprehensions over finite domains, user-defined predicates and more.}


\flatzinc{} is a lower-level language whose purpose is to bridge the
gap between the high-level modeling in \minizinc{} and the need for a
fixed, and easy-to-parse, format that can simplify the implementation
of the input interface of \minizinc{} solvers.
A \minizinc{} model is typically flattened into a \flatzinc{} instance
using the \mzntofzn{} compiler \cite{minizinc_url}, and then solved
with some \minizinc{} tool.




%% file: src/mzn2omt/introduction.tex

We consider the problem of translating \minizinc{} models into \omt{}
problems first.
Similarly to other \minizinc{} solvers, we assume that the \minizinc{} model
is first translated into \flatzinc{} using the \mzntofzn{} standard compiler, as
depicted in Figure~\ref{fig:trans_schema}.
We describe the main aspects of \fzntoomt{}, focusing on the challenges we have
encountered and on the solutions we have adopted.


\begin{figure}[t]
\includegraphics[width=\textwidth]{./data/schema3.pdf}
\caption[Circular Translation Schema]{
\label{fig:trans_schema}
Circular translation schema from \minizinc{} to \smtlib{} and back,
resulting from the composition of \mzntofzn{}, \optimathsat{} and
\omttomzn{}. In this picture, \optimathsat{} acts both as a \flatzinc{}/\omt{}
solver, and also as a \flatzinc{} to \smtlib{} compiler.
}
\end{figure}


%% file: src/mzn2omt/challenges.tex

\paragraph{\flatzinc{} data-types.}
The first challenge is to find a suitable representation of the
data-types supported by \flatzinc{} in \smtlib{}.

One possible choice for modeling the three basic scalar types of \flatzinc{}
--namely {\tt bool}, {\tt int} and {\tt float}-- with \smtlib{} are the
Boolean, bit-vector and floating-point theories respectively.
However, the decision procedures for the bit-vector and floating-point numbers
theories can be significantly more resource demanding than the decision
procedure for the linear arithmetic theory (\lira{}), in particular when
dealing with a substantial amount of arithmetic computations.
For this reason, we have opted to model \flatzinc{} {\tt int}
and {\tt float} data-types with the \smtlib{} integer and
rational types respectively, by default.
For the case in which no substantial linear arithmetic computation is performed,
we also optionally allow for encoding the \flatzinc{} {\tt int} data-type as a
\smtlib{} bit-vector.

For what concerns the two compound types of \flatzinc{}, that is the
{\tt set} and {\tt array} data-types, we have chosen to proceed
as follows.
Given that \optimathsat{} lacks a decision procedure for the theory of finite
sets \cite{sets_url}, we model a set using the Boolean and integer theories,
similarly to what has been done in \cite{bof12}.
The basic idea is to introduce a fresh Boolean variable for each element
in the domain of a set, and use such variable as a placeholder for the
membership of an integer element to the set instance.
Differently from \cite{bof12}, we make an extensive use of cardinality networks
\cite{Asin2011}
to encode constraints over the sets because they are handled more
efficiently, for their nice arc-consistency properties.
No action is required to encode a \flatzinc{} {\tt array} into \smtlib{},
because it is used only as a container for other variables.


\paragraph{Floating-Point precision.}
A consequence of encoding the \flatzinc{} {\tt int} and {\tt float} data-types
with the linear arithmetic theory is that all of our computation is performed
with infinite-precision arithmetic.
This can result in a performance disadvantage wrt. other \minizinc{} solvers
using finite-precision arithmetic, due to the increased cost of each operation,
but it has the benefit of guaranteeing the correctness of the final result of
the computation.%

Currently, the \minizinc{} language does not allow one to express a certain
quantity as an infinite-precision fraction between two constant numbers.
Instead, the \mzntofzn{} compiler computes on-the-fly the result of any division
operation between two constant integers or floating-point numbers applying the
rules corresponding to the type of the operands.
However, there are some instances in which we really need to be able to both
express quantities and perform computation with infinite-precision arithmetic.
One of such situations is to double-check the correctness of the \minizinc{}
models generated by the \omttomzn{} compiler described in
Section~\sref{sec:omt2mzn}
(we have done this for the experimental evaluations in
Section~\sref{sec:omt2mzn_expeval}).
In order to get around this limitation we developed a simple wrapper
around the \mzntofzn{} compiler, called \emzntofzn{}~\cite{emzntofzn_url},
that replaces any fraction among two constant {\tt floats} with a
fresh variable and, after the basic \mzntofzn{} compiler generated
the \flatzinc{} model, the \emzntofzn{} wrapper restores the original
fractional values using the \flatzinc{} constraint {\tt float\_div()}.


\paragraph{\flatzinc{} constraints.}
The \smtlib{} encoding of the majority of \flatzinc{} constraints in \optimathsat{}
follows their definition in the \flatzinc{} Standard, with the exception of
Pseudo-Boolean constraints, which we examine in detail later on.
Several global constraints are also supported in the same way, because the
\omt{}-solver currently lacks ad hoc and efficient decision procedures for
dealing with them.
Constant values and alias variables (e.g. those arising from the definition of
some arrays) are propagated through the formula, so as to keep the set of problem
variables as compact as possible.
Those constraints requiring non-linear arithmetic --like, e.g., trigonometric,
logarithmic and exponential functions-- are currently not supported;
this situation may change soon due to the recent
extension of \mathsatfive{} with a procedure for
it \cite{ac_tocl18}.


\paragraph{Pseudo-Boolean constraints.}
When dealing with Pseudo-Boolean sums of the form $\sum_{i=1}^{i=N}A_i \cdot w_i$,
where $A_i$ is a Boolean variable and $w_i$ is a numerical weight, the \mzntofzn{}
compiler associates a fresh $0$/$1$ integer variable $x_i$ to each $A_i$, and
encodes the sum as $\sum_{i=1}^{i=N}x_i \cdot w_i$.
Notice that the original $A_i$s may not be eliminated from the \flatzinc{} model,
because they typically occur elsewhere in the problem, i.e. as part of a Boolean
formula.
From our own experience, this situation arises frequently, because Pseudo-Boolean
sums are typically used to express cardinality constraints that have a variety of
uses.
As described in \cite{st_tacas17}, one limitation of this naive approach is
that \smt and \omt solvers do not typically handle this encoding 
efficiently.
%
%
%
The main reason is that the pruning power of the conflict clause resulting from a
conflicting assignment is typically limited to one specific Boolean assignment at a
time, meaning that a large number of conflict clauses (possibly exponential) has to
be generated along the search.
As shown in \cite{st_tacas17}, \smt and \omt solvers can benefit from encoding
Pseudo-Boolean constraints with cardinality networks.

\fzntoomt{} goes through some effort in order to recognize Pseudo-Boolean sums
over the integers, and replace the naive encoding with one based on cardinality
networks.
We note that using this technique generally results in a trade-off between
solving time and the overhead of generating cardinality networks prior to starting
the search, especially when dealing with a large number of variables.
%


\paragraph{Multi-objective optimization.}
\fzntoomt{} allows for multiple optimization goals, of heterogeneous type,
being defined within the same \flatzinc{} model.
This is a non-standard extension to the \flatzinc{} format.
Multiple objectives can be solved independently from one another, or combined
into a Lexicographic or Pareto optimization goal.
We refer the reader to \cite{st_jar18} for details on the
input encoding and the solver configuration.


%% file: src/mzn2omt/solutions.tex

\paragraph{Functionality.}
Given a satisfiability or optimization problem encoded in the \flatzinc{}
format, \optimathsat{} can be used in the following ways (Figure~\ref{fig:trans_schema}):
\begin{itemize}
\item to directly solve the problem, optionally enumerating any sub-optimal
      solution found during the search or all possible solutions
      with the same optimal value;
\item to produce an \omt problem encoded with the extended \smtlib{}
      format described in \cite{st_jar18}. This problem can be directly
      solved with \optimathsat{} or, with minor transformations\footnote{To make this step as easy as possible, we collected our scripts
      into a public repository \cite{fzntoomt_url}.}, fed as
      input to other \omt{} solvers such as \bclt{} and \zthree{}.
\end{itemize}


%% file: src/omt2mzn/introduction.tex

In this section, we consider the problem of translating \omt{} formulas,
encoded in the optimization-extended \smtlib{} format of \cite{st_jar18},
into \minizinc{} models.
Hereafter, we describe the main challenges we have faced and the solutions
we have adopted.
Further details about this conversion are available in \cite{omt2mzn}.


%% file: src/omt2mzn/challenges.tex

\paragraph{General Translation Approach.}


\ignoreinshort{
\begin{figure}[t]
    \centering
    \includegraphics[scale=0.45]{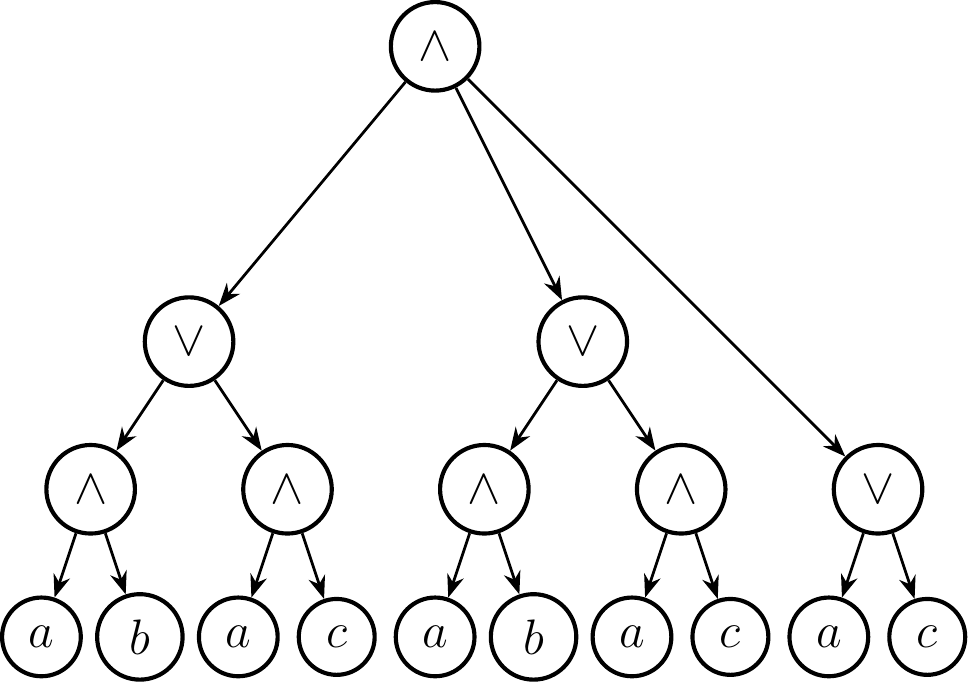}
    \hspace{0.2cm}
    \includegraphics[scale=0.45]{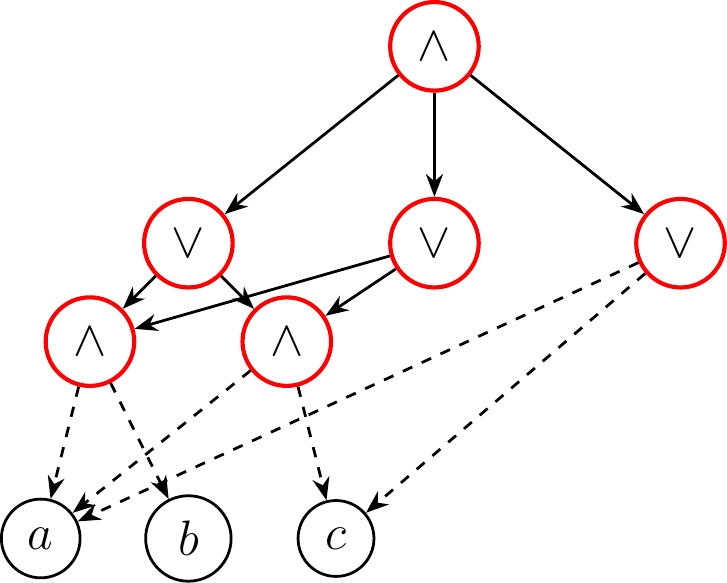}
    \includegraphics[scale=0.45]{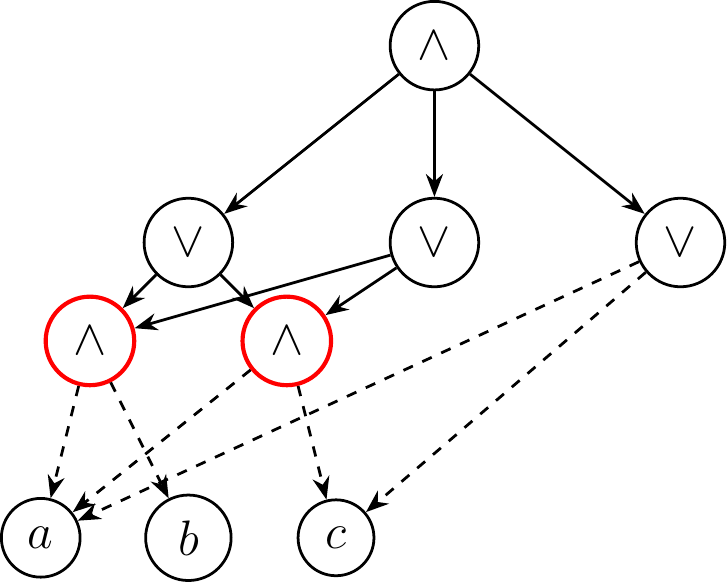}
    \caption[]{
        \label{fig:formula-rap}
             Tree and DAG representation of $((a \land b) \lor (a \land c)) \land ((a \land b) \lor (a \land c)) \land (a \lor c)$.
             The leftmost figure is its tree representation which presents duplication.
             The second and third images are the DAG representation of the same formula.
             The red nodes are the formulae for which a label is created, thus they portray resp.
             the behavior of the \textit{Daggify print} and the \textit{Two-Fathers print}.}
\end{figure}
}


The main challenge is to design an
encoding from \omt{} to \minizinc{} that is {\em correct} (i.e., it
preserves in full the semantics 
of the input \omt{} problems),
{\em effective} (i.e., it produces  as output \minizinc{} models which
are as compact and
easy-to-solve as possible), and  {\em efficient} 
(i.e. it does it with the least consumption of time and memory).
%
%
\ignoreinlong{%
To this extent, one critical design choice is the way in which the
internal representation of the input \omt{} formula 
 is organized and converted in terms of \minizinc{}
primitives.
After a preliminar experimental evaluation we determined that the sweet-spot,
in terms of compactness and easiness to solve of the resulting
\minizinc{} model,
is to adopt what we call ``$\ge$2-father DAG-ification'': a Directed-Acyclic-Graph (DAG) internal representation of
the formula where 
a fresh label is associated to all and only DAG nodes with at least two fathers,
inlining all other nodes
(see \cite{cts_cpaior20_extended} for details).
}
\ignoreinshort{%
For best effectiveness, the internal representation of the input \omt{} formula
adopted by the compiler is that of a Directed Acyclic Graph (DAG),
that represents repeating sub-formula elements with the same node.
A simple, tree-like, traversal of this data-structure can result in some
node --and its entire sub-tree-- being visited multiple times during the
translation of the \omt formula, and hence in a potentially exponential blow-up of
the generated \minizinc{} model (left-most tree in Figure~\ref{fig:formula-rap}).
The converse approach, that is, to assign an explicit label to each node of
the DAG and visit each node exactly once, can suffer from the creation
of a large number of fresh variables that do not exist in the original
problem, even for those nodes that would be traversed only once
(center DAG in Figure~\ref{fig:formula-rap}).
To cope with this problem, we adopted the idea of assigning a label only to
those DAG nodes that have at least two father nodes
(right-most DAG in Figure~\ref{fig:formula-rap}).
This limits the number of artificial variables introduced in the encoding,
and at the same time produces the most compact representation of the
input problem.
In our experiments, the latter idea showed to be essential to produce much
more compact and easy-to-solve problems.
}


\paragraph{Theories Restriction.}

The \smtlib{} standard describes a wide number of \smt theories, most of which
have no direct counterpart in \minizinc{}
due to the few data-types supported (see \sref{sec:background_mzn}).
\ignoreinlong{
Hence, hereafter we deal only with the theory of linear rational and integer
arithmetic, and their combination.
On this regard, we note that even though \omttomzn{} can also
handle the theory of bit-vectors, we do not cover it here because it is not
used in the experimental evaluation in Section~\sref{sec:expeval} (We cover
it in the long version of this paper \cite{cts_cpaior20_extended}).}
\ignoreinshort{
Hence, hereafter we deal only with the theory of linear rational and integer
arithmetic, their combination, and the theory of bit-vectors.}
We leave the handling of other \smt{} theories to future work.


\paragraph{Linear Arithmetic Theory.}

On the surface, encoding linear arithmetic constraints coming from \omt{} in
\minizinc{}, using the {\it int} and {\it float} data-types, looks like a
trivial task.
In reality, this poses several challenges and it is subject to several
limitations, due to a couple of facts.

First,  in \smtlib{} the linear arithmetic theory
requires the capability to perform {\em infinite-precision} computations.
Unfortunately, to the best of our knowledge, no \minizinc{} solver provides
infinite-precision arithmetic reasoning, and the \mzntofzn{} compiler
itself prevents representing arbitrarily-large and arbitrarily-precise
quantities (e.g. the fine-grained decimal weights of the machine
learning application in \cite{teso17}).

Second,  in \omt{} linear arithmetic variables are
not required to be bounded and have
quite often no explicit domain (i.e. they lack a lower-bound, an
upper-bound or both), because it is not necessary for the problem
at hand or it is implied by other constraints.
This is in contrast with \minizinc{}, whereby linear arithmetic variables are
expected to have a finite domain and, when they lack one, their domain appears
to be capped with some solver-dependent pair of values.


These restrictions are currently part of the  \minizinc{} language and the
target application domain, and we do not see any obvious work-around solving
them.
We note that although there exist methods for bounding all
variables in a given LP problem (e.g. \cite{papadimitriou81}),
these have been deemed too impractical at this stage of our investigation.
Nonetheless, we have chosen to translate \smtlib{} linear arithmetic
constraints with a corresponding \minizinc{} encoding based on the
{\it int} and {\it float} data-types.
Although the encoding is not always applicable, it does still allow one
to correctly translate a number of interesting \omt{} problems into \minizinc{},
as witnessed by our experimental evaluation in Section~\sref{sec:omt2mzn_expeval}.
More in detail, the translation is done as follows.
We declare each integer variable as unbounded, and then extend the \minizinc{}
model with the appropriate constraints bounding its domain when the input \omt{}
formula contains any such information.
Our empirical observation is that \minizinc{} models generated in this way are
correctly handed by all \minizinc{} solvers which we have tried, with
the exception of 
\gurobi{}, which returns an ``unsupported'' message.
Floating-Point variables, instead, are always declared with a user-defined
domain. This is because all of the \minizinc{} solvers we have tried, among
those that can handle floating-point constraints, require such information.


\begin{IGNOREINSHORT}
\paragraph{Bit-Vector Theory.}
Dealing with bit-vectors is also challenging.
In essence, an \omt{} bit-vector can be seen as a vector of bits of
fixed size, that can be represented as a Boolean array in \minizinc{}.
However, because bit-vectors are typically used to encode machine
registers and numbers at the binary-level, the bit-vectors theory
provides both signed and unsigned arithmetic operators, as well as
those based on Boolean logic.
Performing arithmetic computation over a Boolean array in \minizinc{}
requires one to define and instantiate of the Boolean circuit encoding
the given operation as a set of constraints.
As far as we can see, this has two main disadvantages.
First, it can result in a massive blow-up of the formula size, due
to the nature of some operations (e.g. multiplication of two bit-vectors).
Second, heavily relying on Boolean reasoning may put \minizinc{}
solvers at a disadvantage when trying to solve the generated formulas.


An alternative solution is to represent \smtlib{} bit-vectors using
the \minizinc{} {\it int} data-type treated as a binary sequence of
bits.
On the one hand, this approach provides the most-efficient handling
of arithmetic operations over bit-vectors (both signed and unsigned).
On the other hand, this encoding can only be applied to bit-vectors
whose width is less than the {\it int} data-type used by the
target \minizinc{} application (i.e. less than $64$ bits).
All things considered, we opted for this approach because it resulted
in the most compact, and easy to handle, encoding of the problem.
\end{IGNOREINSHORT}


\paragraph{Other \omt{} Functionalities.}

Several problems of \omt{} interest require the capability of dealing with
soft-constraints (i.e. Weighted \maxsmt{}) and also with multiple objectives,
that are either considered independent goals or combined in a Lexicographic
or Pareto-like fashion.
To the best of our knowledge, the \minizinc{} standard does not allow for an
explicit encoding of soft-constraints, nor to deal with more than one
objective function at a time.


We encode (weighted) \maxsmt{} problems using a standard Pseudo-Boolean
encoding, such as the one used in \cite{st-ijcar12}.
When dealing with \omt{} problems that contain
$N$
goals $\cost_1, ..., \cost_N$, for $N > 1$,
we use the following approach.
If these objectives are
independent targets,
we generate $N$ \minizinc{} models, each with a different
goal $\cost_i$, and separately solve each model.
If instead the  multiple objectives
belong to a Lexicographic \omt{} problem,
then we generate a unique \minizinc{} model that leverages
the lexicographic-optimization functionality provided by \minisearch{}
\cite{MiniSearch15}. (In all other cases, \minisearch{} is
not used).
We do not have any encoding for dealing with Pareto-optimization, yet.




%% file: src/expeval.tex

In this section we present an extensive empirical evaluation comparing
\omt{} tools with many state-of-the-art \cp{} tools
on \cp{} problems coming from the \minizinc{} challenge (\sref{sec:mzn2omt_expeval}),
and on \omt{} problems coming mostly from formal verification
(\sref{sec:omt2mzn_expeval}). 

The \omt{} solvers under evaluation are \bclt{}, \optimathsat{}
(v. 1.6.0)
and \zthree{} (v. 4.8.5).
These are compared with some of the top-scoring solvers that
participated at recent editions of the \minizinc{} challenge,
including
\choco{} (v. 4.0.4),
\chuffed{},
\gfd{} (v. 1.6.0),
\gecode{} (v. 6.0.1),
\gurobi{} (v. 8.0.1),
\hcsp{} (v. 1.3.0),
\jacop{} (v. 4.5.0),
\izplus{} (v. 3.5.0),
\ortools{} (v. 6.7.4981) and
\picat{} (v. 2.4).
%
\ignoreinshort{\begin{remark}
We could not include \fzntosmt{} \cite{bof10sys,bof12} in our
experimental evaluation because it is not compatible with the 
features of \minizinc{} that have been added since version 2.0.
\end{remark}}

\ignoreinlong{\begin{remark}We could not include \fzntosmt{} \cite{bof10sys,bof12} in our
experimental evaluation because it is not compatible with the 
features of \minizinc{} that have been added since version 2.0.\end{remark}}

We run all these experimental evaluations on two
identical \textit{8-core 2.20Ghz Xeon} machines with $64GB$ of RAM and
running \textit{Ubuntu Linux}.
All the benchmark-sets, the tools and the scripts used to run these
experiments, and some of the plots for the results in
Tables~\ref{tab:mcboth}-\ref{tab:reals} which could not fit into this paper,
can be downloaded from \cite{expeval_url}.

We stress the fact that the goal of these experiments is not to
establish a winner among \omt{}  and \minizinc{} tools; rather,
it is to assess the correctness, effectiveness and efficiency of our
\omt{}-to-\cp{} and \cp{}-to-\omt{} encoders and, more generally, to investigate the
feasibility of solving \minizinc{} problems with \omt{} tools and vice
versa, and to identify the criticalities in terms of efficiency and
correctness in these processes.


%% file: src/mzn2omt/expeval.tex

We consider the benchmark-sets used at the \minizinc{} Challenge of $2016$ (MC16)
and $2019$ (MC19), each comprised by $100$ instances.
For compatibility reasons, the version of \mzntofzn{} used to convert the problems
to the \flatzinc{} format differs between the two benchmark-sets. We use
version $2.2.1$ and $2.3.2$ (with patches) for the problems in MC16 and MC19
respectively.
Due to recent changes in the \flatzinc{} format that affect the benchmarks
in MC19, the version of some \minizinc{} tools differs from what described
in Section~\sref{sec:expeval} (see Table~\ref{tab:mcboth}). In some cases,
we had to download and compile the latest source available for the
tool, i.e. the ``nightly'' version.

We run each \minizinc{} solver with the corresponding directory of global
constraints, and we run each \minizinc{} and \omt{} tool
with the default  options. 
We consider two \omt{} encodings of the original \flatzinc{} problems,
{\sc la} and {\sc bv}. The first encodes the \flatzinc{} \emph{int} type
with the theory of \emph{linear integer arithmetic}, whereas the second
is based on the theory of \emph{bit-vectors}.
We evaluate each \omt{} solver on both \smtlib{} encodings, except for
\bclt{} that has no support for bit-vector optimization.
For uniformity reasons with the other \omt{} solvers,
we evaluate
\optimathsat{} using its \smtlib{} interface only, using thus its
\fzntoomt{} interface as an external tool, like with the other \omt solvers.
We note that the solving time for all \omt{} solvers includes the time
required for translating the formula from the \flatzinc{} to the \smtlib{}
format.
Each solver, either \omt{} or \minizinc{}, is given up to $1200s.$ to solve each problem, not including the
time taken by \mzntofzn{} to flatten it.

We verify the correctness of the results by automatically checking that all
terminating solvers agree on the (possibly optimal) solution and, when this
is not the case, we manually investigate the inconsistency.


\paragraph{Experiment Results.}

 
\input{src/tables/small.tex}

The results of this experiment are shown in Table~\ref{tab:mcboth},
with separate numbers for satisfiability (\textbf{s}) and optimization
(\textbf{o}) instances in each benchmark-set.
Using the experimental data, we separately computed the virtual best
configuration among all \minizinc{} solvers (i.e. {\sc \vbest(\minizinc{})}),
all \omt{} solvers (i.e. {\sc \vbest(\omt{})}),  and also the virtual best
among all tools considered in the experiment (i.e. {\sc \vbest(all)}).
The last two columns in the table list the number of problems solved by
the given configuration in the same amount of time as the {\sc \vbest{}()}
of each group (col. {\em BT1}) and as the {\sc \vbest{}(all)} (col. {\em BT2}).

We start by looking at the \minizinc{} solvers in Table~\ref{tab:mcboth}.
The performance ladder is dominated by \ortoolssat{} and \picatsat{},
closely followed by \gurobi{}, \hcsp{} and \chuffed{} (in MC19).
By looking at column {\em BT1}, we observe that the top-performing \minizinc{}
solvers tend to dominate over all the others.
Looking at the results of the MC19 experiment, we notice a significant
increase in the number of errors with respect to the benchmark-set of
the MC16 edition, as well as a handful of problems solved incorrectly.
In the case of \gurobi{} and \picatsat{}, the \mzntofzn{} compiler
encountered an error over a few instances. As a consequence, the total
number of problems is smaller than $100$ for both tools.
After taking a closer look, we ascribe this phenomenon to the recent changes
in the \minizinc{}/\flatzinc{} format, that has created some minor issues with
some tools that have not been adequately updated.

Looking at the \omt{} tools only, we observe that \zthree{} has leading
performance over the other solvers.
When compared to the \minizinc{} solvers,
the \omt{}
solvers place themselves in the middle of the rank on both
benchmark-sets.
%
Given the fact that
none of the \omt{} solvers has specialized procedures or encodings for dealing with
global constraints,
we consider this an interesting result.

%% file: src/tables/small.tex

\begin{table}[tb]
\scriptsize
\centering
\begin{tabularx}{\textwidth}{|X|rcr|rcr|rcr|rcr|rcr|RCR|rcr|rcr|rcr|}
\cline{14-22}
\multicolumn{13}{c}{} &
\multicolumn{9}{|c|}{\bf terminated} &
\multicolumn{6}{c}{} \\
\cline{2-28}
\multicolumn{1}{c|}{} &
\multicolumn{3}{c|}{\bf inst.}     &
\multicolumn{3}{c|}{\bf timeout}   &
\multicolumn{3}{c|}{\bf error}     &
\multicolumn{3}{c|}{\bf unsup.}    &
\multicolumn{3}{c|}{\bf incor.}    &
\multicolumn{3}{c|}{\bf correct}   &
\multicolumn{3}{c|}{\bf time (s.)} &
\multicolumn{3}{c|}{\bf BT1}       &
\multicolumn{3}{c|}{\bf BT2}       \\
\hline
{\bf tool, configuration \& encoding} &
\scsat{} & | & \scopt{} &
\scsat{} & | & \scopt{} &
\scsat{} & | & \scopt{} &
\scsat{} & | & \scopt{} &
\scsat{} & | & \scopt{} &
\scsat{} & | & \scopt{} &
\scsat{} & | & \scopt{} &
\scsat{} & | & \scopt{} &
\scsat{} & | & \scopt{} \\
\hline
\hline

%
\multicolumn{28}{|c|}{\cellcolor{black!15}\textbf{\minizinc{} Challenge 2016}}\\
\hline

{\sc \picatcp{}} & 15 & | & 85 & 9 &|& 70 & 0 &|& 0 & 0 &|& 0 & 0 &|& 0 & 6 &|& 15 & 2281 &|& 6043 & 0 &|& 0 & 0 &|& 0 \\
{\sc \gfd{}} & 15 & | & 85 & 4 &|& 71 & 1 &|& 3 & 0 &|& 0 & 0 &|& 0 & 10 &|& 11 & 4436 &|& 4220 & 0 &|& 0 & 0 &|& 0 \\
{\sc \choco{}()} & 15 & | & 85 & 3 &|& 50 & 0 &|& 0 & 0 &|& 0 & 0 &|& 0 & 12 &|& 35 & 4256 &|& 11423 & 1 &|& 0 & 1 &|& 0 \\
{\sc \izplus{}()} & 15 & | & 85 & 6 &|& 44 & 0 &|& 0 & 0 &|& 0 & 0 &|& 0 & 9 &|& 41 & 999 &|& 5492 & 3 &|& 4 & 3 &|& 4 \\
{\sc \chuffed{}()} & 15 & | & 85 & 2 &|& 40 & 0 &|& 0 & 5 &|& 0 & 0 &|& 0 & 8 &|& 45 & 635 &|& 4187 & 0 &|& 5 & 0 &|& 5 \\
{\sc \jacop{}()} & 15 & | & 85 & 3 &|& 39 & 0 &|& 0 & 0 &|& 0 & 0 &|& 0 & 12 &|& 46 & 3411 &|& 12825 & 0 &|& 0 & 0 &|& 0 \\
{\sc \gurobi{}()} & 15 & | & 85 & 6 &|& 22 & 0 &|& 0 & 0 &|& 0 & 0 &|& 0 & 9 &|& 63 & 2346 &|& 3037 & 0 &|& 15 & 0 &|& 15 \\
{\sc \hcsp{}()} & 15 & | & 85 & 4 &|& 23 & 0 &|& 0 & 0 &|& 0 & 0 &|& 0 & 11 &|& 62 & 591 &|& 4444 & 0 &|& 11 & 0 &|& 11 \\
{\sc \picatsat{}} & 15 & | & 85 & 1 &|& 26 & 0 &|& 0 & 0 &|& 0 & 0 &|& 0 & {\bf\blue{14}} &|& 59 & 151 &|& 7293 & 10 &|& 1 & 10 &|& 1 \\
{\sc \ortoolssat{}} & 15 & | & 85 & 1 &|& 15 & 0 &|& 0 & 0 &|& 0 & 0 &|& 0 & {\bf\blue{14}} &|& {\bf\blue{70}} & 555 &|& 1338 & 1 &|& 45 & 1 &|& 45 \\
\hline{\sc \vbest{}(\minizinc{})} & 15 & | & 85 & 0 &|& 7 & 0 &|& 0 & 0 &|& 0 & 0 &|& 0 & 15 &|& 78 & 146 &|& 3514 & - &|& - & - &|& - \\
\hline\hline{\sc \optimathsat{}(int)} & 15 & | & 85 & 10 &|& 38 & 0 &|& 0 & 0 &|& 0 & 0 &|& 0 & 5 &|& 47 & 604 &|& 4856 & 1 &|& 20 & 0 &|& 0 \\
{\sc \optimathsat{}(bv)} & 15 & | & 85 & 2 &|& 42 & 0 &|& 0 & 0 &|& 0 & 0 &|& 0 & {\bf\blue{13}} &|& 43 & 3664 &|& 8561 & 11 &|& 2 & 0 &|& 0 \\
{\sc \bclt{}(int)} & 15 & | & 85 & 10 &|& 33 & 0 &|& 0 & 0 &|& 0 & 0 &|& 0 & 5 &|& 52 & 1117 &|& 5998 & 0 &|& 15 & 0 &|& 2 \\
{\sc \zthree{}(int)} & 15 & | & 85 & 10 &|& 32 & 0 &|& 0 & 0 &|& 0 & 0 &|& 0 & 5 &|& 53 & 676 &|& 10424 & 0 &|& 11 & 0 &|& 0 \\
{\sc \zthree{}(bv)} & 15 & | & 85 & 5 &|& 28 & 0 &|& 0 & 0 &|& 0 & 0 &|& 0 & 10 &|& {\bf\blue{57}} & 2938 &|& 11113 & 2 &|& 19 & 0 &|& 0 \\
\hline{\sc \vbest{}(\omt{})} & 15 & | & 85 & 1 &|& 21 & 0 &|& 0 & 0 &|& 0 & 0 &|& 0 & 14 &|& 64 & 3842 &|& 6432 & - &|& - & - &|& - \\
\hline\hline{\sc \vbest{}(all)} & 15 & | & 85 & 0 &|& 7 & 0 &|& 0 & 0 &|& 0 & 0 &|& 0 & 15 &|& 78 & 146 &|& 3514 & - &|& - & - &|& - \\
\hline
\hline
\multicolumn{28}{|c|}{\cellcolor{black!15}\textbf{\minizinc{} Challenge 2019}}\\
\hline

{\sc \picatcp{} [2.7b12]} & 10 & | & 90 & 8 &|& 67 & 0 &|& 11 & 0 &|& 5 & 0 &|& 0 & 2 &|& 7 & 54 &|& 1440 & 1 &|& 0 & 1 &|& 0 \\
{\sc \izplus{}()} & 10 & | & 90 & 5 &|& 71 & 0 &|& 4 & 0 &|& 0 & 0 &|& 0 & 5 &|& 15 & 14 &|& 3077 & 1 &|& 3 & 1 &|& 3 \\
{\sc \gfd{}} & 10 & | & 90 & 5 &|& 64 & 0 &|& 10 & 0 &|& 0 & 0 &|& 0 & 5 &|& 16 & 323 &|& 4010 & 0 &|& 0 & 0 &|& 0 \\
{\sc \choco{}(std)} & 10 & | & 90 & 4 &|& 63 & 0 &|& 5 & 0 &|& 0 & 0 &|& 0 & 6 &|& 22 & 415 &|& 4312 & 0 &|& 0 & 0 &|& 0 \\
{\sc \gecode{}() [6.2.0]} & 10 & | & 90 & 4 &|& 63 & 0 &|& 0 & 0 &|& 0 & 0 &|& 0 & 6 &|& 27 & 420 &|& 5094 & 0 &|& 6 & 0 &|& 6 \\
{\sc \jacop{}() [4.8]} & 10 & | & 90 & 4 &|& 55 & 0 &|& 6 & 0 &|& 0 & 0 &|& 0 & 6 &|& 29 & 260 &|& 5467 & 0 &|& 1 & 0 &|& 1 \\
{\sc \hcsp{}()} & 10 & | & 90 & 0 &|& 47 & 0 &|& 10 & 0 &|& 0 & 2 &|& 2 & 8 &|& 31 & 2 &|& 6408 & 4 &|& 7 & 4 &|& 4 \\
{\sc \chuffed{}() [nightly]} & 10 & | & 90 & 0 &|& 43 & 0 &|& 0 & 5 &|& 10 & 0 &|& 0 & 5 &|& 37 & 1 &|& 4886 & 3 &|& 19 & 3 &|& 19 \\
{\sc \gurobi{}() [8.1.1]} & 10 & | & 80 & 0 &|& 48 & 0 &|& 0 & 0 &|& 0 & 0 &|& 0 & {\bf\blue{10}} &|& 32 & 705 &|& 2895 & 2 &|& 6 & 2 &|& 4 \\
{\sc \picatsat{} [2.7b12]} & 10 & | & 90 & 0 &|& 45 & 0 &|& 5 & 0 &|& 0 & 0 &|& 1 & {\bf\blue{10}} &|& 39 & 275 &|& 9894 & 0 &|& 7 & 0 &|& 5 \\
{\sc \ortoolssat{} [nightly]} & 10 & | & 90 & 5 &|& 42 & 0 &|& 3 & 0 &|& 0 & 0 &|& 0 & 5 &|& {\bf\blue{45}} & 8 &|& 7239 & 0 &|& 13 & 0 &|& 11 \\
\hline{\sc \vbest{}(\minizinc{})} & 10 & | & 90 & 0 &|& 29 & 0 &|& 0 & 0 &|& 0 & 0 &|& 0 & 10 &|& 61 & 9 &|& 5247 & - &|& - & - &|& - \\
\hline\hline{\sc \optimathsat{}(int) [1.6.4.1]} & 10 & | & 90 & 5 &|& 62 & 0 &|& 0 & 0 &|& 0 & 0 &|& 0 & 5 &|& 28 & 4 &|& 3650 & 2 &|& 10 & 0 &|& 0 \\
{\sc \optimathsat{}(bv) [1.6.4.1]} & 10 & | & 90 & 4 &|& 59 & 0 &|& 5 & 0 &|& 0 & 0 &|& 0 & 6 &|& 26 & 484 &|& 7271 & 0 &|& 1 & 0 &|& 0 \\
{\sc \bclt{}(int)} & 10 & | & 90 & 5 &|& 60 & 0 &|& 0 & 0 &|& 0 & 0 &|& 0 & 5 &|& 30 & 6 &|& 3369 & 0 &|& 6 & 0 &|& 5 \\
{\sc \zthree{}(int)} & 10 & | & 90 & 5 &|& 64 & 0 &|& 0 & 0 &|& 0 & 0 &|& 0 & 5 &|& 26 & 4 &|& 5358 & 3 &|& 6 & 0 &|& 1 \\
{\sc \zthree{}(bv)} & 10 & | & 90 & 0 &|& 55 & 0 &|& 2 & 0 &|& 0 & 0 &|& 0 & {\bf\blue{10}} &|& {\bf\blue{33}} & 1629 &|& 7550 & 5 &|& 17 & 0 &|& 3 \\
\hline{\sc \vbest{}(\omt{})} & 10 & | & 90 & 0 &|& 48 & 0 &|& 2 & 0 &|& 0 & 0 &|& 0 & 10 &|& 40 & 1624 &|& 5179 & - &|& - & - &|& - \\
\hline\hline{\sc \vbest{}(all)} & 10 & | & 90 & 0 &|& 29 & 0 &|& 0 & 0 &|& 0 & 0 &|& 0 & 10 &|& 61 & 9 &|& 4919 & - &|& - & - &|& - \\

\hline
\end{tabularx}
\caption[Comparison on the \minizinc{} Challenge formulas]{\label{tab:mcboth}
\minizinc{} Challenge formulas.
The columns list the
total number of instances (inst.),
of timeouts (timeout),
of run-time errors (error),
of unsupported problems (unsup.),
of incorrectly solved instances (incor.),
of correctly solved instances (correct),
the total solving time for all solved instances (time),
the number of instances solved in the shortest time within the same category (BT1)
and those solved in the shortest time considering all tools (BT2).
}
\end{table}


%% file: src/omt2mzn/expeval.tex

In this experimental evaluation we use \omt{} formulas taken from well-known,
publicly available, repositories. 
We characterize these benchmark-sets as follows:

\begin{itemize}
\item {\em SAL} [integers]: $66$ SMT-based Bounded
Model Checking and K-Induction parametric problems created with the
\textsc{SAL} model checker \cite{salurl};

\item {\em SAL} [rationals]: as above, with problems on the rationals;

\item {\em Symba} [rationals]: $2632$ {\em bounded}%
\footnote{
We discarded any {\em unbounded} instance in the original benchmark-set in \cite{li_popl14}.
}
software verification instances derived from a set of C programs used in the
Software Verification Competition of 2013 \cite{li_popl14};

\item {\em Jobshop and Strip Packing} [rationals]: $190$ problems taken from
\cite{DBLP:journals/cce/SawayaG05,st-ijcar12};

\item {\em Machine Learning} [rationals]: $510$ \omt instances
generated with the {\sc pyLMT} tool \ignoreinshort{\cite{pylmt_url}} based on
Machine Learning Modulo Theories \cite{teso17}.
\end{itemize}

\ignore{
The first benchmark-set comes in two versions: one dealing with linear
integer optimization, and the other with linear rational optimization.
All other benchmark-sets deal only with linear rational optimization.
}
\noindent
The first benchmark-set  is on the integers, whereas the other four are
on the rationals.
We stress the fact that all formulas contained in all
benchmark-sets are {\em satisfiable}.


\begin{remark}
Although there exists a repository of multi-objective \omt{}
formulas  (e.g. \cite{li_popl14,st_tacas15}), we
have chosen to not include these in our experimental evaluation.
The reason for this is twofold.
First, such comparison would likely be unfair wrt. CP tools
because 
that the workaround for
dealing with multi-independent \omt{} formulas described in
Section~\sref{sec:omt2mzn_challenges} is not competitive with
the integrated optimization schema provided by \omt{} solvers
\cite{li_popl14,st_tacas15}.
In fact, the experimental evidence in \cite{li_popl14,st_tacas15}
collected on a group of \omt{} solvers indicates that the latter
approach can be an order of magnitude faster than the former one.
Second, the workaround for dealing with lexicographic-optimization
is limited by the fact that \minisearch{} is not fully compatible
with recent versions of \minizinc{}, and it only works with a
restricted set of tools.
%
\end{remark}



We have used the \omttomzn{} tool described in
Section~\sref{sec:omt2mzn} 
to translate each \omt{} formula to the \minizinc{} format.
\omttomzn{} is written in Python and it is built on top of \pysmt{} \cite{PySMT}, a
general-purpose Python library for solving \smt{} problems%
, and it is available at \cite{omt2mzn}.
%
During this step, it has been necessary to impose a finite domain
to any unconstrained \smtlib{} rational variable, because otherwise none of the
\minizinc{} solvers would have been able to deal with them.
We have experimented with two different domains: the largest feasible domain
for floating-point variables of $32$ bits (i.e. $\pm 3.402823e+38$) for the first
two benchmark-sets, and the largest feasible domain for integer variables
(i.e. $\pm 2^{31}$) for the last two.


We consider two \optimathsat{} configurations:
{\sc \optimathsat{}(smt)}, solving the original \omt formulas, and
{\sc \optimathsat{}(fzn)}, executed on the generated \minizinc{}
instances.
The benefits of this choice is two-fold.
First, we can double-check the correctness of such encoding, by comparing
the optimum models generated in the two cases.
Second, we can verify whether there is any performance loss caused by
the encoding of the formula.


Only four of the \minizinc{} solvers listed in Section~\sref{sec:expeval} support
floating-point reasoning. This limited the number of tools that could be
used with some \omt{} benchmark-sets.
The running-time of each \minizinc{} solver reported in these experiments
(including {\sc \optimathsat{}(fzn)}) is comprehensive of the time taken
by the \mzntofzn{} compiler, because the latter can sometime solve the
input formulas on its own.
The overall timeout is set to $600s$.


Notice that the optimal solutions found by {\sc \optimathsat{}(smt)} have
been previously independently verified with a third-party \smt{} tool
as reported in previous publications \cite{st_cav15,st_tacas15,st_tacas17}.%
%
\footnote{
For every OMT problem \tuple{\vi,\cost} s.t. {\sc \optimathsat{}(smt)} returns a
minimum value $min$ for $\cost$ on the formula $\vi$, we say $min$ is correct
iff $\vi\wedge(\cost=min)$ is 
satisfiable and $\vi\wedge(\cost<min)$ is unsatisfiable. (Dual for maximization.)
}
Therefore, we verify the correctness of the results found by any other
configuration by comparing them with those found by {\sc \optimathsat{}(smt)},
and otherwise mark the result as ``unverified''.


\ignoreinlong{
\input{src/tables/omt_merged.tex}
}
\ignoreinshort{
\input{src/tables/omt_int.tex}

}


\paragraph{Experimental Results over the Integers.}

In this experiment, we evaluate the {\em SAL (over integers)}
benchmark-set. 
The results are collected in Table~\ref{tab:int}.


%
We notice first that {\sc \optimathsat{}(fzn)} always produces correct
results and it shows comparable
performances in terms on number of problems solved wrt. the baseline {\sc
  \optimathsat{}(smt)}, solving even 4 problems more. 
(We conjecture that the latter fact should be attributed to the limited,
but effective, deduction capabilities of the \mzntofzn{} compiler, that may
have helped \optimathsat{} in solving the input formulas.)
This suggests that, at least on problems on the integers,
 \omttomzn{} is efficient and effective and does not affect
 correctness. 



In general, \minizinc{} solvers do not seem to deal efficiently with this benchmark-set.
Some tools have experienced some internal error (e.g. dumped-core,
segmentation fault), some others have been killed to to a high memory
consumption (over $32$GB), whereas the majority of the remaining tools
had a timeout.

We explain this behavior with the fact that the given benchmark set is
characterized by the presence of a heavy Boolean structure combined
with arithmetical constraints, which
requires the efficient combination of strong Boolean-reasoning 
capabilities (e.g., efficiently 
handling chains of unit propagations) with 
strong arithmetical-solving\&optimization capabilities, which is a
typical feature of \omt{} solvers.
\ignore{
E.g., we conjecture that the fact that \picatsat{} is built on top of
a \sat{} engine may explain why it appears to perform better than the
other \minizinc{} solvers on this experiment.
}

\ignore{%
We explain this behavior with the fact that the given benchmark set is
characterized by the  presence of a heavy Boolean structure, which
requires strong Boolean-reasoning capabilities (e.g., efficiently
handling long chains of unit propagations).
E.g., we conjecture that the fact that \picatsat{} is built on top of
a \sat{} engine may explain why it appears to perform better than the
other \minizinc{} solvers on this experiment.%
}

None of the input formulas was initially supported by \gurobi{}. After
restricting the bound of every integer variable to $\pm 10^6$,
\textsc{\gurobi{}(l)} was able to solve $3$ instances within the
timeout.
Among the \minizinc{} solvers, the best result is obtained by \picatsat{}, that
solved $4$ problems out of $66$.


\ignoreinshort{
\input{src/tables/omt_reals.tex}

}


\paragraph{Experimental Results Over the Rationals.}

We consider first the first three benchmark-sets over the rationals:
SAL over rationals, Symba, JobShop\&Strip-Packing.
\ignoreinlong{(Separate tables for the four 
benchmarks are reported in the extended version of this paper \cite{cts_cpaior20_extended}.)}
Of all \minizinc{} solvers we have tried, only three are able to deal
with floating-point constraints.
The results are shown in Table~\ref{tab:reals}.
Since each of the input formulas is satisfiable, we consider a
result incorrect either when it is equal to \unsatres, or when the
relative error $\Delta$ exceeds $10^{-6}$, s.t.:
$
    \Delta\defas\frac{|o_{smt}-o_{fzn}|}{|o_{smt}|},
$
$o_{smt}$ and $o_{fzn}$ being the optimal value found by
{\sc \optimathsat{}(smt)} and the optimal value found by the
\minizinc{} solver under test respectively. (Recall that the former
was previously checked to be correct.)


Similarly to the previous experiment on the integers, 
{\sc \optimathsat{}(fzn)} always produces correct results,
and display comparable performance wrt. {\sc \optimathsat{}(omt)}
in terms of number of instances being solved, solving somewhat
fewer problems.
This is not the case of the other three \minizinc{} solvers.
Among these, \gecode{} experienced a timeout on the majority of the formulas
being considered, \gmip{} returned mostly incorrect answers, whereas
\gurobi{} seems to have the best performance, in particular on the
third benchmark-set.


We attribute the large number of incorrect results returned by all three
\minizinc{} solvers to the fact that these tools use finite-precision
floating-point arithmetic internally.
The incorrect behavior of some of these solvers (e.g. \gurobi{}) can also
be partially explained with the large domain of floating-point variables in
these problems. However, given the nature of these input instances,
it was not possible for us to assign a smaller domain to each variable
in the problem {\em a priori}.


\smallskip
We analyze separately the results for the last benchmark-set reported in
Table~\ref{tab:reals}.
The peculiar aspect of the Machine Learning benchmark-set
\cite{teso17} is that it is
characterized by Pseudo-Boolean sums over rational weights, and by
very fine-grained
rational values%
\footnote{For example, 
$\frac{1799972218749879}{2251799813685248}$
\ignoreinshort{, which expands to $0.79934824037669782725856748584192...$, }
is a sample weight value from problems in \cite{teso17}.}.
%
%
%
Unfortunately, these fine-grained rational values are rounded by the
standard \mzntofzn{} compiler, which causes the incorrect results even of 
{\sc \optimathsat{}(fzn)} in Table~\ref{tab:reals}, despite the fact
that \optimathsat{} uses infinite-precision arithmetic.

In order to overcome this issue, we leverage the \emzntofzn{} compiler
described in Section~\sref{sec:mzn2omt_challenges} so that the original
fractional values are preserved in the resulting \flatzinc model, and
show that with this approach \optimathsat{} does not produce incorrect
results any longer
(configuration \optimfzne{} in Table~\ref{tab:reals}),
solving correctly $152$ problems more than \optimfzn{}.


\ignore{
We observe that, in general, {\sc \optimathsat{}(fzn+E)} solves fewer
formulas than {\sc \optimathsat{}(smt)} on this benchmark-set.
{\ptchange{}
This performance gap is due to the structural changes in
the formula after using the \omttomzn{} and the \mzntofzn{}
compilers. This prevents {\sc \optimathsat{}(fzn+E)} to
use some efficient solving techniques when dealing
with the same problem given to {\sc \optimathsat{}(smt)}.
}
}

Overall, since there are at least $237$ formulas affected by the above
issue with the \mzntofzn{} compiler, we avoid
an in-depth discussion of the results obtained by the other
\minizinc{} solvers. However, at a first glance the situation
does not seem to differ from the other benchmark-sets over the
rationals.


%% file: src/tables/omt_merged.tex

\begin{table}[t]
\scriptsize
\centering
\begin{tabularx}{\textwidth}{|X|r|r|r|r|r|x|r|r|r|r|}
\cline{6-10}
\multicolumn{5}{c|}{} & \multicolumn{5}{c|}{\bf terminated}\\
\hline
\textbf{tool \& configuration} &
\textbf{inst.}  &
\textbf{timeout} &
\textbf{tool-er.} &
\textbf{unsupp.} &
\textbf{incor.} &
\textbf{correct} &
\textbf{tot. time (s.)} &
\textbf{avg. time (s.)} &
\textbf{med. time (s.)} \\
\hline
    \gurobi{}()             & 66 & 0  & 0  & 66 &0 & 0 & 0 & 0.00 & 0.00 \\
    \gfd{}                  & 66 & 0  & 66 & 0  &0 & 0 & 0 & 0.00 & 0.00 \\
    \izplus{}()             & 66 & 0  & 66 & 0  &0 & 0 & 0 & 0.00 & 0.00 \\
    \jacop{}()              & 66 & 0  & 66 & 0  &0 & 0 & 0 & 0.00 & 0.00 \\
    \chuffed{}()            & 66 & 19 & 47 & 0  &0 & 0 & 0 & 0.00 & 0.00 \\
    \ortoolssat             & 66 & 57 & 9  & 0  &0 & 0 & 0 & 0.00 & 0.00 \\
    \choco{}()              & 66 & 66 & 0  & 0  &0 & 0 & 0 & 0.00 & 0.00 \\
    \hcsp{}()               & 66 & 66 & 0  & 0  &0 & 0 & 0 & 0.00 & 0.00 \\
    \picatcp{}              & 66 & 66 & 0  & 0  &0 & 0 & 0 & 0.00 & 0.00 \\
    \gecode{}()             & 66 & 66 & 0  & 0  &0 & 0 & 0 & 0.00 & 0.00 \\
    \sc \gurobi{}(l)        & 66 & 63 & 0  & 0  &0 & 3 & 166  & 55.49 & 52.44  \\
    \picatsat{}             & 66 & 62 & 0  & 0  &0 & 4 & 1667 & 416.85 & 467.09  \\
    \hline
    \vbestpar{\minizinc{}}    & 66 & 62 & 0  & 0  &0 & 4 & 718 & 179.51 & 78.54   \\
\hline
\hline
    \optimfzn               & 66 & 18 & 0 & 0   &0 & \textcolor{blue}{\textbf{48}} & 7113 & 148.20 & 70.52 \\
\hline
    \vbestpar{fzn}          & 66  & 18  & 0 & 0 &0 & 48 & 7113 & 148.20 & 70.52 \\
\hline
\hline
    \optimsmt               & 66 & 22 & 0  & 0  &0 & 44& 2657 & 60.41 & 18.72 \\
\hline
    \vbestpar{all}          & 66  & 16  & 0 & 0 &0 & 50 & 5037 & 100.75 & 25.13 \\
\hline
\end{tabularx}
\caption{\label{tab:int}
SAL over integers.
%
A \satres result is marked as {\em correct} when the objective
value matches the reference solution provided by \optimsmt{}
(when run without a timeout),
as {\em incorrect} otherwise.
}




\begin{tabularx}{\textwidth}{|X|r|r|r|r|x|y|r|r|r||r|r|r|r|r|r|}
\cline{5-16}
\multicolumn{4}{c|}{} & \multicolumn{6}{c||}{\bf terminated} & \multicolumn{6}{c|}{\bf incorrect results}\\
\hline
\textbf{tool \& configuration} &
\rotatebox{90}{\textbf{instances}} &
\rotatebox{90}{\textbf{timeout}} &
\rotatebox{90}{\textbf{tool-errors}} &
\rotatebox{90}{\textbf{incorrect}} &
\rotatebox{90}{\textbf{verified}} &
\rotatebox{90}{\textbf{unverified}} &
\rotatebox{90}{\textbf{tot. time (s.)}} &
\rotatebox{90}{\textbf{avg. time (s.)}} &
\rotatebox{90}{\textbf{med. time (s.)}} &
\rotatebox{90}{\textbf{unsat}} &
\rotatebox{90}{$ \Delta \geq 10^{-6} $} &
\rotatebox{90}{$ \Delta \geq 10^{-3} $} &
\rotatebox{90}{$ \Delta \geq 10^{-1} $} &
\rotatebox{90}{$ \Delta \geq 10^{0} $} &
\rotatebox{90}{$ \Delta \geq 10^{1} $} \\
\hline\hline
\multicolumn{16}{|c|}{\cellcolor{black!15}\textbf{SAL, Symba, Jobshop and Strippacking}} \\
\hline
    \gecode{}() & 2888 & 2733 & 0 & 0 & 155 & 0 & 10800 & 69.68 & 18.63 & 0 & 0 & 0 & 0 & 0 & 0 \\
    \gmip{} & 2888 & 10 & 0 & 2855 & 0 & 23 & 317 & 13.79 & 11.21 & 2765 & 90 & 90 & 86 & 39 & 0 \\
    \gurobi{}() & 2888 & 48 & 0 & 2728 & 104 & 8 & 3961 & 35.37 & 2.14 & 2684 & 44 & 32 & 32 & 1 & 0 \\
\hline
    \vbestpar{\minizinc{}} & 2888 & 0 & 0 & 2628 & 237 & 23 & 13801 & 53.08 & 6.97 & - & - & - & - & - & - \\
\hline\hline
    \optimfzn{} & 2888 & 31 & 0 & 0 & 2854 & 3 & 22320 & 7.81 & 0.40 & 0 & 0 & 0 & 0 & 0 & 0 \\
\hline
    \vbestpar{fzn} & 2888 & 0 & 0 & 11 & 2854 & 23 & 20674 & 7.19 & 0.40 & - & - & - & - & - & - \\
\hline\hline
    \optimsmt{} & 2888 & 23 & 0 & 0 & \textcolor{blue}{\textbf{2865}} & 0 & 15676 & 5.47 & 0.08 & 0 & - & - & - & - & - \\
\hline\hline
    \vbestpar{all}            &  2888 & 0 & 0 & 0 & 2865 & 23 & 15183 & 5.26 & 0.08 & - & - & - & - & - & - \\
\hline\hline
\multicolumn{16}{|c|}{\cellcolor{black!15}\textbf{Machine Learning}} \\
\hline
    \gecode{}() & 510 & 322 & 0 & 164 & 24 & 0 & 11 & 0.44 & 0.43 & 147 & 17 & 17 & 2 & 0 & 0 \\
    \gmip{} & 510 & 108 & 0 & 400 & 2 & 0 & 225 & 112.47 & 112.47 & 400 & 0 & 0 & 0 & 0 & 0 \\
    \gurobi{}() & 510 & 9 & 0 & 472 & 28 & 1 & 201 & 6.92 & 3.17 & 468 & 4 & 4 & 2 & 0 & 0 \\
\hline
    \vbestpar{\minizinc{}} & 510 & 9 & 0 & 464 & 36 & 1 & 383 & 10.34 & 0.46 & - & - & - & - & - & - \\
\hline\hline
    \optimfzn & 510 & 7 & 0 & 237 & 263 & 3 & 2797 & 10.52 & 2.21 & 177 & 60 & 59 & 0 & 0 & 0 \\
    \optimfzne & 510 & 92 & 0 & 0 & 415 & 3 & 1197 & 2.86 & 2.03 & 0 & 0 & 0 & 0 & 0 & 0 \\
\hline
    \vbestpar{fzn} & 510 & 7 & 0 & 83 & 417 & 3 & 1366 & 3.25 & 2.03 & - & - & - & - & - & - \\
\hline\hline
    \optimsmt & 510 & 10 & 0 & 0 &\textcolor{blue}{\textbf{500}} & 0 & 5766 & 11.53 & 12.15 & 0 & - & - & - & - & - \\
\hline\hline
    \vbestpar{all} & 510 & 7 & 0 & 0 & 500 & 3 & 2290 & 4.55 & 2.05 & - & - & - & - & - & - \\
\hline
\end{tabularx}
\caption{\label{tab:reals}
\omt Problems defined over the rationals.
%
%
A \satres result is marked as {\em correct} when the objective
value matches the reference solution provided by \optimsmt{}
with an absolute error $\Delta < 10^{-6}$.
A result is marked as {\em unverified} when we have
no reference solution and {\em incorrect} if neither
of the previous two conditions apply.
%
}

\end{table}


%% file: src/tables/omt_int.tex

\begin{table}[t]
\scriptsize
\begin{tabularx}{\textwidth}{|X|r|r|r|r|r|x|r|r|r|r|}
\cline{6-10}
\multicolumn{5}{c|}{} & \multicolumn{5}{c|}{\bf terminated}\\
\hline
\textbf{tool \& configuration} &
\textbf{inst.}  &
\textbf{timeout} &
\textbf{tool-er.} &
\textbf{unsupp.} &
\textbf{incor.} &
\textbf{correct} &
\textbf{tot. time (s.)} &
\textbf{avg. time (s.)} &
\textbf{med. time (s.)} \\
\hline
    \gurobi{}()             & 66 & 0  & 0  & 66 &0 & 0 & 0 & 0.00 & 0.00 \\
    \gfd{}                  & 66 & 0  & 66 & 0  &0 & 0 & 0 & 0.00 & 0.00 \\
    \izplus{}()             & 66 & 0  & 66 & 0  &0 & 0 & 0 & 0.00 & 0.00 \\
    \jacop{}()              & 66 & 0  & 66 & 0  &0 & 0 & 0 & 0.00 & 0.00 \\
    \chuffed{}()            & 66 & 19 & 47 & 0  &0 & 0 & 0 & 0.00 & 0.00 \\
    \ortoolssat             & 66 & 57 & 9  & 0  &0 & 0 & 0 & 0.00 & 0.00 \\
    \choco{}()              & 66 & 66 & 0  & 0  &0 & 0 & 0 & 0.00 & 0.00 \\
    \hcsp{}()               & 66 & 66 & 0  & 0  &0 & 0 & 0 & 0.00 & 0.00 \\
    \picatcp{}              & 66 & 66 & 0  & 0  &0 & 0 & 0 & 0.00 & 0.00 \\
    \gecode{}()             & 66 & 66 & 0  & 0  &0 & 0 & 0 & 0.00 & 0.00 \\
    \sc \gurobi{}(l)        & 66 & 63 & 0  & 0  &0 & 3 & 166  & 55.49 & 52.44  \\
    \picatsat{}             & 66 & 62 & 0  & 0  &0 & 4 & 1667 & 416.85 & 467.09  \\
    \hline
    \vbestpar{\minizinc{}}    & 66 & 62 & 0  & 0  &0 & 4 & 718 & 179.51 & 78.54   \\
\hline
\hline
    \optimfzn               & 66 & 18 & 0 & 0   &0 & \textcolor{blue}{\textbf{48}} & 7113 & 148.20 & 70.52 \\
\hline
    \vbestpar{fzn}          & 66  & 18  & 0 & 0 &0 & 48 & 7113 & 148.20 & 70.52 \\
\hline
\hline
    \optimsmt               & 66 & 22 & 0  & 0  &0 & 44& 2657 & 60.41 & 18.72 \\
\hline
    \vbestpar{all}          & 66  & 16  & 0 & 0 &0 & 50 & 5037 & 100.75 & 25.13 \\
\hline
\end{tabularx}
\caption{\label{tab:int}
SAL over integers.
The configuration \textsc{\gurobi{}(l)} is given in input
a modified version of the input problem in which every
integer variable has been forcibly bounded by $\pm 10^6$.
The columns list the
total number of instances (inst.),
the number of timeouts (timeout),
of tool errors (tool-er.),
of unsupported problems (unsupp.),
of instances terminating within the timeout (Terminated),
of instances with an incorrect solution (incor.),
of instances with a correct solution (correct) and
the total/average/median solving time for all solved instances (tot./avg./med. time).
A \satres result is marked as {\em correct} when the objective
value matches the reference solution provided by \optimsmt{}
(when run without a timeout),
as {\em incorrect} otherwise.
}
\end{table}


%% file: src/tables/omt_reals.tex

\begin{table}[pt]
\scriptsize
\begin{tabularx}{\textwidth}{|X|r|r|r|r|x|y|r|r|r||r|r|r|r|r|r|}
\cline{5-16}
\multicolumn{4}{c|}{} & \multicolumn{6}{c||}{\bf terminated} & \multicolumn{6}{c|}{\bf incorrect results}\\
\hline
\textbf{tool \& configuration} &
\rotatebox{90}{\textbf{instances}} &
\rotatebox{90}{\textbf{timeout}} &
\rotatebox{90}{\textbf{tool-errors}} &
\rotatebox{90}{\textbf{incorrect}} &
\rotatebox{90}{\textbf{verified}} &
\rotatebox{90}{\textbf{unverified}} &
\rotatebox{90}{\textbf{tot. time (s.)}} &
\rotatebox{90}{\textbf{avg. time (s.)}} &
\rotatebox{90}{\textbf{med. time (s.)}} &
\rotatebox{90}{\textbf{unsat}} &
\rotatebox{90}{$ \Delta \geq 10^{-6} $} &
\rotatebox{90}{$ \Delta \geq 10^{-3} $} &
\rotatebox{90}{$ \Delta \geq 10^{-1} $} &
\rotatebox{90}{$ \Delta \geq 10^{0} $} &
\rotatebox{90}{$ \Delta \geq 10^{1} $} \\

\hline\hline
\multicolumn{16}{|c|}{\cellcolor{black!15}\textbf{SAL over rationals}} \\
\hline
    \gecode(){} & 66 & 66 & 0 & 0 & 0 & 0 & 0 & 0.00 & 0.00 & 0 & 0 & 0 & 0 & 0 & 0 \\
    \gmip{} & 66 & 0 & 0 & 54 & 0 & 12 & 305 & 25.43 & 25.81 & 54 & 0 & 0 & 0 & 0 & 0 \\
    \gurobi(){} & 66 & 40 & 0 & 26 & 0 & 0 & 0 & 0.00 & 0.00 & 15 & 11 & 11 & 11 & 0 & 0 \\
\hline
    \vbestpar{\minizinc{}} & 66 & 0 & 0 & 54 & 0 & 12 & 305 & 25.43 & 25.81 &- &- &- &- &- &- \\
\hline\hline
    \optimfzn & 66 & 16 & 0 & 0 & 47 & 3 & 6952 & 139.03 & 66.18 & 0 & 0 & 0 & 0 & 0 & 0 \\
\hline
    \vbestpar{fzn} & 66 & 0 & 0 & 7 & 47 & 12 & 6096 & 103.32 & 31.67 & - & - & - & - & - & -\\
\hline\hline
    \optimsmt & 66 & 12 & 0 & 0 & \textcolor{blue}{\textbf{54}} & 0 & 4907 & 90.87 & 29.31 &0&-&-&-&-&- \\
\hline\hline
    \vbestpar{all} & 66 & 0 & 0 & 0 & 54 & 12 & 5212 & 78.97 & 27.17 & - & - & - & - &- & -\\
\hline\hline
\multicolumn{16}{|c|}{\cellcolor{black!15}\textbf{Software Verification (Symba)}} \\
\hline
    \gecode{}() & 2632 & 2477 & 0 & 0 & 155 & 0 & 10800 & 69.68 & 18.63 & 0 & 0 & 0 & 0 & 0 & 0 \\
    \gmip{} & 2632 & 0 & 0 & 2632 & 0 & 0 & 0 & 2632 & 0.00 & 0.00 & 0 & 0 & 0 & 0 & 0 \\
    \gurobi{}() & 2632 & 0 & 0 & 2596 & 36 & 0 & 103 & 2595 &2.87 & 2.12 & 1 & 1 & 1 & 1 & 0 \\
\hline
    \vbestpar{\minizinc{}} & 2632 & 0 & 0 & 2463 & 169 & 0 & 10402 & 61.55 & 9.41 &- &- &- &- &- &-\\
\hline\hline
    \optimfzn & 2632 & 0 & 0 & 0 & 2632 & 0 & 1366 & 0.52 & 0.39 & 0 & 0 & 0 & 0 & 0 & 0\\
\hline
    \vbestpar{fzn} & 2632 & 0 & 0 & 0 & 2632 & 0 & 1366 & 0.52 & 0.39 &- &- &- &- &- &-\\
\hline\hline
    \optimsmt & 2632 & 0 & 0 & 0 & 2632 & 0 & \textcolor{blue}{\textbf{284}} & 0.11 & 0.08 & 0 & - & - & - & - & -\\
\hline\hline
    \vbestpar{all} & 2632 & 0 & 0 & 0 & 2632 & 0 & 283 & 0.11 & 0.08 &- &- &- &- &- &-\\
\hline\hline
\multicolumn{16}{|c|}{\cellcolor{black!15}\textbf{JobShop + Strippacking}} \\
\hline
    \gecode{}() & 190 & 190 & 0 & 0 & 0 & 0 & 0 & 0.00 & 0.00 & 0 & 0 & 0 & 0 & 0 &0\\
    \gmip{} & 190 & 10 & 0 & 169 & 0 & 11 & 12 & 1.10 & 1.21 & 79 & 90 & 90 & 86 & 39 &0 \\
    \gurobi{}() & 190 & 8 & 0 & 106 & 68 & 8 & 3858 & 50.76 & 2.76 & 74 & 32 & 20 & 20 & 0 &0 \\
\hline
    \vbestpar{\minizinc{}} & 190 & 0 & 0 & 111 & 68 & 11 & 3093 & 39.16 & 2.17 &- &- &- &- &- &-\\
\hline\hline
    \optimfzn & 190 & 15 & 0 & 0 & 175 & 0 & 14002 & 80.01 & 18.85 & 0 & 0 & 0 & 0 & 0 & 0\\
\hline
    \vbestpar{fzn} & 190 & 0 & 0 & 4 & 175 & 11 & 13211 & 71.03 & 13.58 & - & - & - & - & - & - \\
\hline
\hline
    \optimsmt & 190 & 11 & 0 & 0 & \textcolor{blue}{\textbf{179}} & 0 & 10484 & 58.57 & 14.66 & 0 & - & - & - & - & -\\
\hline\hline
    \vbestpar{all} & 190 & 0 & 0 & 0 & 179 & 11 & 9687 & 50.98 & 9.97 &- &- &- &- &- &-\\
\hline\hline
\multicolumn{16}{|c|}{\cellcolor{black!15}\textbf{Machine Learning}} \\
\hline
    \gecode{}() & 510 & 322 & 0 & 164 & 24 & 0 & 11 & 0.44 & 0.43 & 147 & 17 & 17 & 2 & 0 & 0 \\
    \gmip{} & 510 & 108 & 0 & 400 & 2 & 0 & 225 & 112.47 & 112.47 & 400 & 0 & 0 & 0 & 0 & 0 \\
    \gurobi{}() & 510 & 9 & 0 & 472 & 28 & 1 & 201 & 6.92 & 3.17 & 468 & 4 & 4 & 2 & 0 & 0 \\
\hline
    \vbestpar{\minizinc{}} & 510 & 9 & 0 & 464 & 36 & 1 & 383 & 10.34 & 0.46 & - & - & - & - & - & - \\
\hline\hline
    \optimfzn & 510 & 7 & 0 & 237 & 263 & 3 & 2797 & 10.52 & 2.21 & 177 & 60 & 59 & 0 & 0 & 0 \\
    \optimfzne & 510 & 92 & 0 & 0 & 415 & 3 & 1197 & 2.86 & 2.03 & 0 & 0 & 0 & 0 & 0 & 0 \\
\hline
    \vbestpar{fzn} & 510 & 7 & 0 & 83 & 417 & 3 & 1366 & 3.25 & 2.03 & - & - & - & - & - & - \\
\hline\hline
    \optimsmt & 510 & 10 & 0 & 0 &\textcolor{blue}{\textbf{500}} & 0 & 5766 & 11.53 & 12.15 & 0 & - & - & - & - & - \\
\hline\hline
    \vbestpar{all} & 510 & 7 & 0 & 0 & 500 & 3 & 2290 & 4.55 & 2.05 & - & - & - & - & - & - \\
\hline
\end{tabularx}
\caption{\label{tab:reals}
\omt Problems defined over the rationals.
%
%
A \satres result is marked as {\em correct} when the objective
value matches the reference solution provided by \optimsmt{}
with an absolute error $\Delta < 10^{-6}$.
A result is marked as {\em unverified} when we have
no reference solution.
A result is marked as {\em incorrect} if neither
of the previous two conditions apply.
}
\end{table}


%% file: src/discussion_features.tex

%
On the whole, from our experiments, \omt{} tools appear to be still at some
 disadvantage when dealing
with \minizinc{} problems wrt. specific tools, and vice versa.
%


On the one hand, \omt{} solvers seem to be penalized by their lack of efficient
ad hoc decision procedures for dealing with global constraints.
Moreover, the approach taken by the \mzntofzn{} compiler, that creates lots
of alias Boolean, integer and floating-point variables for dealing
with Pseudo-Boolean constraints, is particularly challenging to
deal with efficiently by an \omt solver.

On the other hand, \minizinc{} solvers seem to suffer with
problems needing an arithmetic-reasoning component
combined with heavy  Boolean-reasoning component.
Even more importantly, the lack of infinite-precision linear
arithmetic procedures causes a number of incorrect results
when dealing with \omt{} problems over the rationals.
Both of these points need to be addressed in order to deal with
the vast number of Formal Verification and Model Checking
applications in the \smt{}/\omt{} domain.


%% file: src/conclusions.tex

In this paper we have taken a first step forward towards bridging
the \minizinc{} and the \omt{} communities.
The ultimate goal is to obtain a correct, effective and efficient 
fully-automated system for translating
problems from one community to the other, so as to extend the application
domain of both communities.
With our experimental evaluation, we have identified some criticalities
that need to be addressed by each community in order to
solidify this union.


We plan to push this investigation forward as follows.
In the short term, we plan to address the inefficient handling of
Pseudo-Boolean constraints over the rationals revealed by the
experimental evaluation in Section~\sref{sec:omt2mzn_expeval}.
In order to deal with those \flatzinc{} constraints that
require non-linear arithmetic, we envisage an opportunity to either
extend \optimathsat{} with proper handling of the non-linear arithmetic
theory \cite{ac_tocl18} or to experiment with an encoding based on
the floating-point theory \cite{ts_cade19}.
This objective goes hand in hand with the extension of \omttomzn{}
to deal with other \smt{} theories.
In the long term, \omt{} solving may also benefit from adopting
efficient ad hoc decision procedures for frequently used global
constraints.
Finally, we plan to broaden the scope of our investigation and
include other \omt{} solvers in our study.

